\documentclass[11pt]{article}
\usepackage[margin=1in]{geometry}  
\usepackage{graphics,graphicx,color}
\usepackage{amssymb,amsfonts,amsthm,amsmath,mathrsfs}
\usepackage{multirow, booktabs}
\usepackage{subcaption}
\usepackage{algorithm}
\usepackage{algpseudocode}
\usepackage[toc,page]{appendix}
\usepackage{authblk}
\usepackage{natbib}
\usepackage[colorlinks=true, linkcolor=blue, citecolor=blue]{hyperref}
\usepackage{url}
\usepackage{bm}
\usepackage{bbm}
\usepackage[super]{nth}
\usepackage{setspace} 
\usepackage{xr}
\newtheorem{example}{Example}

\newtheorem{theorem}{Theorem}
\newtheorem{proposition}{Proposition}

\theoremstyle{remark}

\theoremstyle{plain}
\newcommand{\diag}{\mbox{diag}}

\newcommand{\change}[1]{\textcolor{black}{#1}}

\begin{document}

\title{Bayesian $D$-Optimal Design of Experiments with Quantitative and Qualitative Responses}

\author{Lulu Kang$^1$, Xinwei Deng$^2$, Ran Jin$^2$}
\affil{$^1$Illinois Institute of Technology, Chicago, IL, $^2$Virginia Tech, Blacksburg, VA.}
\date{}

\maketitle

\begin{abstract}
Systems with both quantitative and qualitative responses are widely encountered in many applications.
Design of experiment methods are needed when experiments are conducted to study such systems.
Classic experimental design methods are unsuitable here because they often focus on one type of response.
In this paper, we develop a Bayesian $D$-optimal design method for experiments with one continuous and one binary response.
Both noninformative and conjugate informative prior distributions on the unknown parameters are considered. The proposed design criterion has meaningful interpretations regarding the $D$-optimality for the models for both types of responses.
An efficient point-exchange search algorithm is developed to construct the local $D$-optimal designs for given parameter values.
Global $D$-optimal designs are obtained by accumulating the frequencies of the design points in local $D$-optimal designs, where the parameters are sampled from the prior distributions.
The performances of the proposed methods are evaluated through two examples.
\end{abstract}

\section{Introduction}\label{sec:intro}

In many applications, both quantitative and qualitative responses are often collected for evaluating the quality of the system.
Often, the two types of responses are mutually dependent. 
We call such a system with both types of quality responses quantitative-qualitative system.
Such systems are widely encountered in practice \citep{kang2022generative,kang2021auxillary,kang2018bayesian}.
In \cite{kang2018bayesian}, the authors studied an experiment of the lapping stage of the wafer manufacturing process. 
The qualitative response is the conformity of the site total indicator reading (STIR) of the wafer, which has two possible outcomes: whether or not the STIR of a wafer is within the tolerance. 
The quantitative response is the total thickness variation (TTV) of the wafer. 
\cite{kang2021auxillary} focused on the birth records and examined the mutual dependency of birth weight and preterm birth. 
The birth weight of an infant is a quantitative outcome and the preterm birth is a binary indicator of whether an infant is born before 36 gestational weeks.
The two types of outcomes are correlated as an infant is usually underweight if the infant is born preterm. 
In \cite{kang2022generative}, two case studies of quantitative-qualitative systems from material sciences and gene expressions are illustrated. 
In the gene expression study, the qualitative response has three possible outcomes: healthy individuals, patients with Crohn's disease, and patients with Ulcerative colitis. 

This work is motivated by a study of an etching process in a wafer manufacturing process.
In the production of silicon wafers, the silicon ingot is sliced into wafers in fine geometry parameters.
Inevitably, this step leaves scratches on the wafers' surface.
An etching process is used to improve the surface finish, during which the wafers are submerged in the container of etchant for chemical reaction.
The quality of the wafers after etching is measured by two response variables: the total thickness variation of the wafer (TTV) and the binary judgment that whether the wafer has cloudy stains in its appearance.
The two responses measure the quality from different but connected aspects.
There is a hidden dependency between the continuous TTV and binary judgment of stains.
To improve the etching quality, expensive experiments are to be carried out to reveal this hidden dependency and to model the quality-process relationship.
Therefore, an ideal experimental design for continuous and binary responses should be able to extract such useful information with economic run size.

The classic design of experiments methods mainly focus on experiments with a single continuous response. 
There have been various methods developed for a single discrete response too, including \cite{sitter1995optimal, woods2006designs, russell2008, woods2010robust, woods2012blocked}.
For multiple responses, \cite{draper1966design} proposed the seminal work for continuous responses, \cite{denman2011design} developed a design method for bivariate binary responses modeled by Copula functions. 
In the case of mixed types of responses, the literature is very scarce. 
A naive design method is to combine the two designs that are separately constructed for each type of response.
However, such a naive strategy could be reasonable for one type of response but problematic for the other by ignoring the dependency between the types of responses, as shown in Example 1.

\begin{figure}[h]
\begin{center}
\includegraphics[scale=0.53]{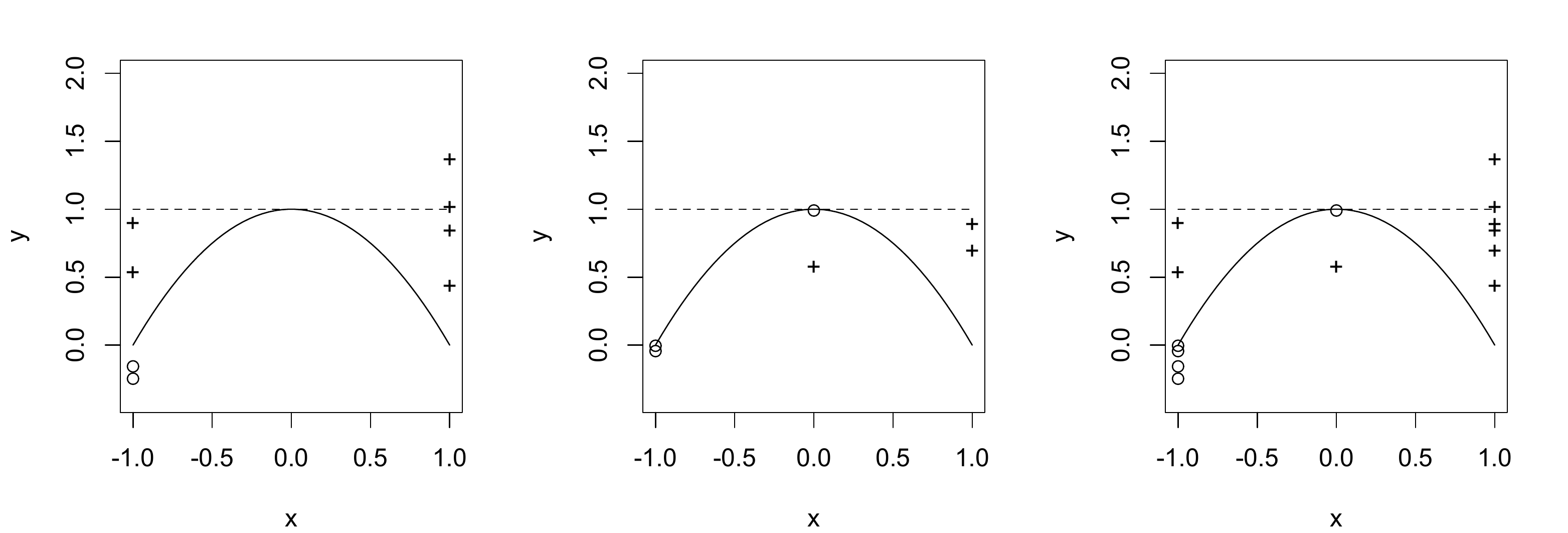}
\caption{(a) Observations from design for $Z$; (b) Observations from design for $Y$; (c) Observations from the combined design. Dashed line ``- - -'' denotes $\mathbb{E}(Y|Z=1)=1$; solid line ``---'' denotes $\mathbb{E}(Y|Z=0)=1-x^2$; point ``+'' denotes $(x_i, y_i)$ with $z_i=1$; point ``o'' denotes $(x_i,y_i)$ with $z_i=0$. \label{fig:eg1}}
\end{center}
\end{figure}

\begin{example}
Denote by $Y$ and $Z$ a continuous response and a binary response, respectively.
Assume that the true model of the binary response $Z$ is $\mathbb{E}(Z|x)=\pi(x)=\exp(1+x)/(1+\exp(1+x))$. 
The true model of $Y$ is related to $Z$ in the form $Y|Z=z \sim N(1-(1-z)x^2, 0.3^2)$.
Thus, $\mathbb{E}(Y|Z=1)=1$ and $\mathbb{E}(Y|Z=0)=1-x^2$. 
Using the naive design method, a 14-point design is constructed, which consists of an 8-point local $D$-optimal design for the model of $Z$ with $\log(\pi(x)/(1-\pi(x))=\eta_0+\eta_1x$, and a 6-point $D$-optimal design for linear regression with a quadratic model of $x$. 
Given the design, we generate the responses from the true models of $Y$ and $Z$. Figure \ref{fig:eg1} (a)-(c) show the data $(x_i, y_i, z_i)$ from the 8-point, 6-point and their combined 14-point design, respectively.
\end{example}

In this example, there is a strong dependency between the two responses since the true underlying models of $\mathbb{E}(Y|Z)$ are different when $Z=1$ and $Z=0$.
In both designs for a single response shown in Figure \ref{fig:eg1} (a) and (b), the design points are balanced and reasonably distributed for the targeted response.
However, since there are no $Y$ observations for $Z=0$ at $x=1.0$ shown in Figure \ref{fig:eg1} (c), the quadratic model for $Y|Z=0$ is not estimable.
Clearly, the combined design is not suitable here.
Note that this problem is not caused by outliers, since all the points for $Z=1$ (with ``+'') are varied around $Y=1$ and the points for $Z=0$ (with ``o'') are around $Y=1-x^2$.
In fact $P(Z=0|x=1)=0.12$, which is relatively small.
Thus it is less likely to observe $Y$ with $Z=0$ at $x=1.0$.
A simple solution is to add more replications at $x=1.0$, but it is not clear how many replications should be sufficient.
It becomes more difficult to spot a direct solution when the experiments get more complicated.

Such experiments call for new experimental design methods to account for both continuous and binary responses.
Note that under the experimental design framework, the linear model is often considered for modeling the continuous response, and the generalized linear model (GLM) is often considered for modeling the qualitative response.
A joint model must be developed to incorporate both types of responses. 
Compared to the classic design methods for linear models or GLMs, design for the joint model is more challenging due to the following aspects.
First, the design criterion for the joint model is more complicated, as the joint model is more complicated than the separate ones. 
Second, experimental design for the GLM itself is more difficult than that for the linear model, which is naturally inherited by the design for the joint model.
Third, efficient design construction algorithms are needed to handle the complexity of the design criterion based on the joint model.
\cite{kang2019bayesian} proposed an $A$-optimal design for the experiments with both quantitative and qualitative responses. 
The $A$-optimality was derived under a Bayesian framework proposed in \cite{kang2018bayesian}. 
Although \cite{kang2019bayesian} addressed the three challenges to a degree, the $A$-optimality is not a commonly used criterion. 
More importantly, only informative prior is considered, which circumvented some difficulties brought by noninformative prior of the parameters. 

In this paper, we choose the most commonly used $D$-optimal design criterion and propose a novel Bayesian design method for the continuous and binary responses.
The proposed method considers both cases of noninformative priors and informative priors. 
With the noninformative priors, the Bayesian framework is equivalent to the frequentist approach. 
In this case, we also establish some regularity conditions on the experimental run sizes.
With the informative priors,  we develop the $D$-optimal design using conjugate priors.
The derived design criterion has meaningful interpretations in terms of the $D$-optimality criteria for the models of both continuous and binary responses.
Moreover, we develop an efficient point-exchange algorithm to construct the proposed designs.
The construction algorithm can be applied to more general settings other than factorial designs.

The rest of the paper is organized as follows.
Section \ref{sec:QQ_MnD} reviews the general Bayesian quantitative-qualitative (QQ) model and the optimal design criterion.
The Bayesian $D$-optimal design criterion is derived using noninformative prior distributions in Section \ref{sec:noninformative}.
In Section \ref{sec:conjugate}, the design criterion is derived with conjugate informative priors.
Efficient algorithms for constructing optimal designs are elaborated in Section \ref{sec:alg}.
One artificial example and the etching experimental design are shown in Section \ref{sec:egs}.
Section \ref{sec:end} concludes this paper with some discussions.

\section{General Bayesian QQ Model and Design}\label{sec:QQ_MnD}

We first review the general Bayesian QQ model introduced in \cite{kang2018bayesian} and focus on the scenario that $Y$ is a continuous response and $Z$ is a binary response.
The input \change{variable} $\bm x = (x_{1}, \ldots, x_{p})'\in \mathbb{R}^p$ contains $p$ dimensions.
Denote the data as $(\bm x_{i}, y_{i}, z_{i}), i = 1, \ldots, n$,  where $y_{i} \in \mathbb{R}$ and $z_{i} \in  \{ 0, 1 \}$.
The vectors $\bm y=(y_i, \ldots, y_n)'$ and $\bm z=(z_1, \ldots, z_n)'$ are the vectors of response observations.
To jointly model the continuous response $Y$ and the binary response $Z$ given $\bm x$, consider the joint probability of $Y|Z$ and $Z$.
The conditional model on $Y|Z$ is assumed to be a linear regression model, while the model of $Z$ is a logistic regression model.
Specifically, we consider joint modeling of $Y$ and $Z$ as follows,
\begin{align}\label{eq:model-Z}
Z = \left \{
\begin{array}{cl}
1, & \textrm{ with probability } \pi(\bm x) \\
0, & \textrm{ with probability } 1-\pi(\bm x)
\end{array}
\right.
\textrm{ with } \pi(\bm x,\bm \eta) =  \frac{\exp(\bm f(\bm x)' \bm \eta)}{1+\exp(\bm f(\bm x)' \bm \eta)},
\end{align}
where $\bm f(\bm x)=\left(f_1(\bm x), \ldots, f_q(\bm x)\right)$ contains $q$ modeling effects including the intercept, the main, interaction and quadratic effects, etc., and $\bm \eta = (\eta_{1}, \ldots, \eta_q)'$ is a vector of parameter coefficients.
Conditioning on $Z=z$, the quantitative variable $Y$ has the distribution
\begin{equation} \label{eq:model-Y}
Y|Z=z \sim N(\mu_0+z\bm f(\bm x) ' \bm \beta^{(1)}+ (1-z)\bm f(\bm x)'\bm \beta^{(2)}, \sigma^{2}),
\end{equation}
where $\bm \beta^{(i)} = (\beta^{(i)}_{1}, \ldots, \beta^{(i)}_q)', i =1, 2$ are the corresponding coefficients of the model effects.
The parameter $\mu_0$ is the mean and $\sigma^2$ is the noise variance.
The above \change{conditional model} \eqref{eq:model-Y} indicates that $Y|Z=1 \sim N(\mu_0+\bm f(x)'\bm\beta^{(1)}, \sigma^{2})$ and $Y|Z=0 \sim N(\mu_0+\bm f(\bm x)'\bm \beta^{(2)},\sigma^{2})$.
We assume the same variance $\sigma^2$ for the two conditional distributions of $Y|Z=1$ and $0$.
The design method developed in the paper can be easily adapted to the case with different variances.

The association between the two responses $Y$ and $Z$ is represented using the conditional model $Y|Z$.
When the two linear models for $Y|Z=0$ and $Y|Z=1$ are different, i.e., $\bm \beta^{(1)} \ne \bm \beta^{(2)}$, then it is important to take account of the influence of the qualitative response $Z$ when modeling the quantitative response $Y$.
Let \change{$\bm X=(\bm x_1, \ldots, \bm x_n)'$} be the $n\times p$ design matrix with $\bm x_i$ as the $i$th design point.
Based on the CB model, we can express the sampling distributions as
\begin{align}
&\bm y|\bm z, \bm \beta^{(1)}, \bm \beta^{(2)}, \mu_0, \sigma^2, \bm X \sim N(\mu_0{\bf 1}+\bm V_1\bm F\bm \beta^{(1)}+\bm V_2\bm F\bm \beta^{(2)}, \sigma^2\bm I_n),\\
&\bm z|\bm \eta, \bm X\sim \textrm{Bernoulli}(\pi(\bm x_i,\bm \eta))\textrm{ for } i=1,\ldots, n, \textrm{ and }  \\\nonumber
&p(\bm z|\bm \eta, \bm X)\propto \exp \left\{\sum_{i=1}^n\left(z_i \bm f(\bm x_i)'\bm \eta- \log (1+e^{\bm f(\bm x_i)'\bm \eta})\right)\right\},
\end{align}
where $p(\cdot)$ denotes a general density function.
Here $\bm V_1=\diag\{z_1,\ldots, z_n\}$ is a diagonal matrix, $\bm I_n$ is the $n\times n$ identity matrix and $\bm V_2=\bm I_n-\bm V_1$, $\bm F$ is the model matrix with the $i$th row as $\bm f(\bm x_i)'$, and $\bf 1$ is a vector of ones.

Denote $p(\bm \beta^{(1)})$, $p(\bm \beta^{(2)})$, and $p(\bm \eta)$ as the prior distributions of the parameters $\bm \beta^{(1)}$, $\bm \beta^{(2)}$, and $\bm \eta$.
Note that we focus on the estimation accuracy of the three groups of parameters.
The mean $\mu_0$ and variance $\sigma^2$ are considered nuisance parameters and thus excluded from the optimal design criterion.
In this work, we assume that the priors of $\bm \beta^{(1)}$, $\bm \beta^{(2)}$, and $\bm \eta$ are independent.
Under this assumption, the conditional posterior distribution of $\bm \eta$, $\bm \beta^{(1)}$, and  $\bm \beta^{(2)}$ are also independent as explained in Sections 3 and 4.
Under the Bayesian framework, the conditional posterior distribution of the parameters $(\bm \beta^{(1)}, \bm \beta^{(2)}, \bm \eta)$ can be derived as
\begin{align}
\label{eq:gnr_poster}
p(\bm \beta^{(1)}, \bm \beta^{(2)}, \bm \eta|\bm y, \bm z, \mu_0, \sigma^2, \bm X)
&\propto p(\bm y|\bm z, \bm \beta^{(1)}, \bm \beta^{(2)}, \mu_0, \sigma^2, \bm X)p(\bm \beta^{(1)})p(\bm \beta^{(2)})p(\bm z|\bm \eta, \bm X)p(\bm \eta)
\end{align}
Using \eqref{eq:gnr_poster} we develop the general Bayesian optimal design criterion.
Let $\psi$ be a criterion function on the conditional posterior distribution of the parameters.
For example, it can be the Shannon information (or equivalently, the Kullback-Leibler distance), $A$/$I$-optimality \citep{fedorov1972theory}, or other design criteria.
However, $\psi(\cdot)$ cannot be directly used as the final optimal design criterion because its value depends on the random parameters $(\bm \beta^{(1)}, \bm \beta^{(2)}, \bm \eta)$ and the experimental outputs $(\bm y, \bm z)$ that are not yet observed. 
The randomness of $(\bm \beta^{(1)}, \bm \beta^{(2)}, \bm \eta)$ can be removed by calculating the mean of $\psi$ with respect to these parameters. 
The uncertainty of $(\bm y, \bm z)$ can be removed by calculating the mean $\mathbb{E}(\mathbb{E}(\psi|\bm y, \bm z))$. 
Therefore, the general Bayesian optimal design criterion on the design matrix $\bm X$ is
\begin{align}
\label{eq:gen_crit}
\Psi(\bm X|\mu_0,\sigma^2)=\int p(\bm y, \bm z|\mu_0,\sigma^2, \bm X)&\times\left(\int \psi(p(\bm \beta^{(1)}, \bm \beta^{(2)}, \bm \eta|\bm y, \bm z,\mu_0, \sigma^2, \bm X))\times\right.\\\nonumber
&\left.p(\bm \beta^{(1)}, \bm \beta^{(2)}, \bm \eta|\bm y, \bm z,\mu_0, \sigma^2, \bm X)d\bm \beta^{(1)}d\bm\beta^{(2)}d\bm \eta \right) d\bm yd\bm z.
\end{align}
It is well-known that the Bayesian $D$-optimal design is equivalent to the Shannon information criterion \citep{chaloner1995bayesian}, omitting the constant terms to $\bm X$.
The criterion function $\psi(\cdot)$ of Shannon information is $\log(\cdot)$.
Next, we develop the Bayesian $D$-optimal design criteria \eqref{eq:gen_crit} under different prior distributions.

\section{Optimal Design under Noninformative Priors}\label{sec:noninformative}

When lacking domain knowledge or proper historical data, experimenters often favor the frequentist approach as no priors need to be specified.
The frequentist approach can be seen as the Bayesian approach using noninformative priors.
In this section, we derive the optimal design criterion and the regularity conditions for noninformative priors.

\subsection{Design Criterion}\label{subsec:noninfo_D}

Assume the non-informative priors $p(\bm \beta^{(i)})\propto 1$ for $i=1, 2$ and $p(\bm \eta)\propto 1$.
The conditional posterior distribution in \eqref{eq:gnr_poster} is the same as the joint distribution of the data.
It can be further factorized into
\[
p(\bm \beta^{(1)}, \bm \beta^{(2)}, \bm \eta|\bm y, \bm z, \mu_0, \sigma^2, \bm X)\propto p(\bm \eta|\bm z, \bm X)\prod_{i=1}^2p(\bm \beta^{(i)}|\bm y, \bm z, \mu_0, \sigma^2, \bm X),
\]
with the posterior distributions
\begin{align}
\label{eq:noninfo_post_beta}
&\bm \beta^{(i)}|\bm y, \bm z,\mu_0,\sigma^2, \bm X\sim N\left((\bm F'\bm V_i\bm F)^{-1}\bm F'\bm V_i(\bm y-\mu_0{\bf 1}), \sigma^2(\bm F'\bm V_i\bm F)^{-1}\right)\textrm{  for } i=1, 2,\\
\label{eq:noninfo_post_eta}
&p(\bm \eta|\bm z, \bm X)\propto \exp \left\{\sum_{i=1}^n\left(z_i \bm f(\bm x_i)'\bm \eta- \log (1+e^{\bm f(\bm x_i)'\bm \eta})\right)\right\}.
\end{align}
Conditioning on $\bm z$, the posterior distributions of $(\bm \beta^{1},\bm \beta^{2})$ and $\bm \eta$ are independent. 
Note that the noninformative prior $p(\bm \eta)\propto 1$ is proper because it leads to proper posterior $p(\bm \eta|\bm z, \bm X)$.
Under the noninformative priors, the Bayesian estimation is identical to the frequentist estimation.

Using the posterior distributions \eqref{eq:noninfo_post_beta}--\eqref{eq:noninfo_post_eta} and the criterion function $\psi(\cdot)=\log(\cdot)$ in the general Bayesian optimal design criterion \eqref{eq:gen_crit}, we obtain the Bayesian $D$-optimal design criterion \eqref{eq:noninfo_Dopt_exact}.
\begin{align}
\label{eq:noninfo_Dopt_exact}
\Psi(\bm X|\mu_0,\sigma^2)&=\mathbb{E}_{\bm z,\bm \eta}\left\{ \log(p(\bm\eta|\bm z, \bm X))\right\}\\\nonumber
&+\frac{1}{2}\sum_{i=1}^2\mathbb{E}_{\bm \eta}\mathbb{E}_{\bm z|\bm \eta}\left\{\log\det\{(\bm F'\bm V_i\bm F)\}\right\}+\textrm{constants}.
\end{align}
The derivation is in Appendix \ref{sec:proofs}.
The first additive term in \eqref{eq:noninfo_Dopt_exact} is exactly the Bayesian $D$-optimal design criterion for GLMs.
Unfortunately, its exact integration is not tractable.
The common approach in experimental design for GLMs is to use a normal approximation for the posterior distribution $p(\bm \eta|\bm z, \bm X)$ \citep{chaloner1995bayesian, khuri2006design}. Such an approximation leads to
\begin{equation}
\label{eq:noninfo-approx1}
\mathbb{E}_{\bm z, \bm \eta}\left\{ \log(p(\bm\eta|\bm z, \bm X))\right\}\approx \mathbb{E}_{\bm \eta}\{\log \det \bm I(\bm \eta|\bm X)\}+\textrm{constant},
\end{equation}
where $\bm I(\bm \eta|\bm X)$ is the Fisher information matrix.
We can easily show that
\[
\bm I(\bm \eta|\bm X)=-\mathbb{E}_{\bm z}\left(\frac{\partial^{2} l(\bm z,\bm \eta|\bm X)}{\partial \bm \eta \partial \bm \eta^{T}}\right) = \sum_{i=1}^{n} \bm f(\bm x_{i}) \bm f(\bm x_{i})' \pi(\bm x_{i}, \bm \eta) (1- \pi(\bm x_{i}, \bm \eta)) = \bm F' \bm W_0 \bm F,
\]
where $\bm W_0$ is a diagonal weight matrix
\[\bm W_0 = \diag\{\pi(\bm x_{1}, \bm \eta) (1-\pi(\bm x_{1}, \bm \eta)), \ldots, \pi(\bm x_{n}, \bm \eta) (1-\pi(\bm x_{n}, \bm \eta))\}.\]
Omitting the irrelevant constant, we approximate the exact criterion $\Psi(\bm X|\mu_0, \sigma^2)$ in \eqref{eq:noninfo_Dopt_exact} as follows.
\begin{align}\label{eq: approx-Psi}
\Psi(\bm X|\mu_0, \sigma^2)\approx \mathbb{E}_{\bm \eta}\{\log\det(\bm F'\bm W_0\bm F)\}+\frac{1}{2}\sum_{i=1}^2\mathbb{E}_{\bm \eta}\mathbb{E}_{\bm z|\bm \eta}\left\{\log\det(\bm F'\bm V_i\bm F)\right\}.
\end{align}

To construct the optimal design, we consider maximizing the approximated $\Psi(\bm X|\mu_0, \sigma^2)$ in \eqref{eq: approx-Psi}.
But this is not trivial, because it involves the expectation on $Z_i$'s in the second additive term.
To overcome this challenge, \cite{hainy2013approximate} constructed optimal designs by simulating samples from the joint distribution of responses and the unknown parameters.
But this method can be computationally expensive for even slightly larger dimensions of experimental factors.
Instead of simulating $Z_i$'s, we derive the following Theorem \ref{the:noninfo-d-criterion} that gives a tractable upper bound $Q(\bm X)$.
Thus we propose using the upper bound $Q(\bm X)$ as the optimal criterion.
\begin{theorem}\label{the:noninfo-d-criterion}
Assume that the matrices $\bm F'\bm W_0\bm F$, $\bm F'\bm V_1\bm F$, and $\bm F'\bm V_2\bm F$ are all nonsingular.
Omitting the irrelevant constant, an upper bound of the approximated $\Psi(\bm X|\mu_0, \sigma^2)$ is
\begin{equation}\label{eq:lower}
Q(\bm X)=\mathbb{E}_{\bm \eta}\left\{\log\det(\bm F'\bm W_0\bm F)+\frac{1}{2}\sum_{i=1}^2\log\det(\bm F'\bm W_i\bm F)\right\},
\end{equation}
where $\bm W_1=\diag\{\pi(\bm x_1,\bm \eta), \ldots, \pi(\bm x_n, \bm \eta)\}$ and $\bm W_2=\bm I_n-\bm W_1$.
\end{theorem}
The proof of Theorem \ref{the:noninfo-d-criterion} is in Appendix \ref{sec:proofs}.
Note that Theorem \ref{the:noninfo-d-criterion} requires that $\bm F'\bm W_i\bm F$ for $i=0, 1, 2$ are all nonsingular.
Obviously $\bm W_0=\bm W_1\bm W_2$. It is easy to see that $\bm F'\bm W_0\bm F$ is nonsingular if and only if both $\bm F'\bm W_1\bm F$ and $\bm F'\bm W_2\bm F$ are nonsingular.

The matrices $\bm F'\bm V_1\bm F$ and $\bm F'\bm V_2\bm F$ involve the responses $Z_i$'s that are not yet observed at the experimental design stage.
We can only choose the experimental run size and the design points to avoid the singularity problem with a larger probability for given values of $\bm \eta$.
Once the run size is chosen, the design points can be optimally arranged by maximizing $Q(\bm X)$.
The weight matrix $\bm W_1$ (or $\bm W_2$) gives more weight to the feasible design points that are more likely to lead to $Z=1$ (or $Z=0$) observations so that the parameters $\bm \beta^{(1)}$ (or $\bm \beta^{(2)}$) of the linear model $Y|Z=1$ (or $Y|Z=0$) are more likely to be estimable.
Next, we introduce some regularity conditions on the run size and number of replications to alleviate the singularity problem.

\subsection{Regularity Conditions}\label{subsec:noninfo_reg}
Let $m$ be the number of distinct design points in the design matrix $\bm X$, $n_i$ be the number of repeated point $\bm x_i$ in $\bm X$ for $i=1, \ldots, m$.
Thus $n=\sum_{i=1}^m n_i$ and $n_i\geq 1$ for $i=1, \ldots, m$.
First, it is necessary that $m\geq q$ for the linear regression model to be estimable under the noninformative priors.
The if-and-only-if condition for $\bm F'\bm W_0\bm F$ to be nonsingular is that $\textrm{rank}(\bm F'\bm W_0\bm F)\geq q$.
If $m\geq q$ and $\pi(\bm x_i, \bm \eta)\in (0, 1)$ for $i=1, \ldots, m$, then $\bm F'\bm W_i\bm F$ for $i=0,1,2$ are all nonsingular and thus positive definite.
To make sure $\pi(\bm x_i, \bm \eta)\in (0, 1)$ for $i=1, \ldots, m$, it is sufficient to assume that $\bm \eta$ is finitely bounded.
This condition is typically used for the frequentist $D$-optimal design for GLMs.
For instance, \cite{woods2006designs} suggested using the centroids of the finite bounded space of $\bm \eta$ to develop the local $D$-optimal design for GLMs.
\cite{dror2006robust} clustered different local $D$-optimal designs with $\bm \eta$ randomly sampled from its bounded space.

The if-and-only-if condition for $\bm F'\bm V_1\bm F$ and $\bm F'\bm V_2\bm F$ being nonsingular is
\begin{equation}
\label{eq:iffcond}
\sum_{i=1}^m I(\sum_{j=1}^{n_i}Z_{ij}>0)\geq q \quad\textrm{and} \quad \sum_{i=1}^m I(\sum_{j=1}^{n_i}Z_{ij}<n_i)\geq q,
\end{equation}
where $Z_{ij}$ is the $j$th random binary response at the unique design point $\bm x_i$ and $I(\cdot)$ is the indicator function.
In the following, we discuss how to choose sample sizes $n_i$ and $n$ under two scenarios (i) $m=q$ and  (ii) $m>q$.

\begin{proposition}\label{prop:cond1}
Assume $m=q$. Both $\bm F'\bm V_1\bm F$ and $\bm F'\bm V_2\bm F$ are nonsingular if and only if $I(0<\sum_{j=1}^{n_i} Z_{ij}<n_i)=1$ for $i=1, 2,\ldots,m$.
For any given $\kappa\in (0, 1)$, a sufficient condition on $n_i$ for $\Pr(0<\sum_{j=1}^{n_i} Z_{ij}<n_i)\geq \kappa$ is
\begin{equation}
\label{eq:regcond1_s}
n_i\geq 1+\left\lceil\frac{\log(1-\kappa)}{\log\left(\max\left\{\pi(\bm x_i, \bm \eta), 1-\pi(\bm x_i, \bm \eta)\right\}\right)}\right\rceil \quad\textrm{for }i=1, 2, \ldots, m,
\end{equation}
and a necessary condition is
\begin{equation}
\label{eq:regcond1_n}
n_i\geq \left\lceil\frac{2\log\left(\frac{1-\kappa}{2}\right)}{\log\pi(\bm x_i,\bm \eta)+\log(1-\pi(\bm x_i, \bm \eta))}\right\rceil  \quad\textrm{for }i=1, 2, \ldots, m.
\end{equation}
\end{proposition}

\begin{proposition}\label{prop:cond2}
Assume $m>q$.
To make both $\bm F'\bm V_1\bm F $ and $\bm F'\bm V_2\bm F$ nonsingular with large probability, or equivalently\change{,}
\[
\mathbb{E}\left\{\sum_{i=1}^m I(\sum_{j=1}^{n_i}Z_{ij}>0)\right\}\geq q \quad\textrm{and} \quad \mathbb{E}\left\{\sum_{i=1}^m I(\sum_{j=1}^{n_i}Z_{ij}<n_i)\right\}\geq q,
\]
\begin{itemize}
\item [(i)] a sufficient condition is
\begin{equation}
\label{eq:regcond2_suf}
n_0\geq \max\left\lceil\left\{1, \frac{\log(1-q/m)}{\log(1-\pi_{\min})}, \frac{\log(1-q/m)}{\log\pi_{\max}}\right\}\right\rceil,
\end{equation}
which is the same as
\begin{equation}
\label{eq:regcond2_stg}
n \geq \left\lceil m\cdot \max\left\{1, \frac{\log(1-q/m)}{\log(1-\pi_{\min})}, \frac{\log(1-q/m)}{\log\pi_{\max}}\right\}\right\rceil, \end{equation}
where $n_0=\min\{n_1,\ldots, n_m\}$, $\pi_{\min}=\min_{i=1}^m\pi(\bm x_i, \bm \eta)$, $\pi_{\max}=\max_{i=1}^m\pi(\bm x_i, \bm \eta)$ and $\bm x_i$'s are the unique design points;
\item [(ii)] a necessary condition is
\begin{equation}
\label{eq:regcond2_weak}
n_0 \geq\left\lceil \max\left\{1, \frac{\log(1-q/m)}{\log(1-\pi_{\max})}, \frac{\log(1-q/m)}{\log\pi_{\min}}\right\}\right\rceil 
\end{equation}
which is the same as
\begin{equation}
\label{eq:regcond2_weak_all}
n \geq\left\lceil  m \cdot \max\left\{1, \frac{\log(1-q/m)}{\log(1-\pi_{\max})}, \frac{\log(1-q/m)}{\log\pi_{\min}}\right\}\right\rceil.\end{equation}
\end{itemize}
\end{proposition}

Proposition \ref{prop:cond1} gives a sufficient condition on the lower bound of $n_i$ when saturated design ($m=q$) is used.
Under the sufficient condition, the nonsingularity of $\bm F\bm V_1\bm F$ and $\bm F\bm V_2\bm F$ holds with a probability larger than $\kappa^m$.
For Example 1 in Section \ref{sec:intro}, suppose that if the possible values of $x$ can only be $-1, 0, 1$, then $m=q=3$.
Let $\bm \eta=(1,1)'$. 
If $\kappa=0.5$, then the numbers of replications for $x=-1, 0, 1$ need to satisfy $n_1\geq 2$, $n_2\geq 4$, and $n_3\geq 7$\change{,} respectively.
\change{If $\kappa=0.9$, then $n_1\geq 5$, $n_2\geq 9$, and $n_3\geq 20$.}
Proposition \ref{prop:cond1} is useful in Step 1 to construct the initial design in Algorithm \ref{alg:global} in Section \ref{sec:alg}.

Proposition \ref{prop:cond2} provides one sufficient condition and one necessary condition when $m>q$ on the smallest number of replications and the overall run size.
But these conditions only examine the nonsingularity of the two matrices with large probability, which is weaker than Proposition \ref{prop:cond1}.
For given $\bm \eta$ value, Algorithm \ref{alg:global} in Section \ref{subsec:local} can return the local $D$-optimal design.
Proposition \ref{prop:cond2} can be useful to check the local $D$-optimal design, as $\pi_{\min}$ and $\pi_{\max}$ depend on $\bm \eta$.
Take the artificial example in Section \ref{subsec:toy} for instance.
The local $D$-optimal design for $\rho=0$ (Table \ref{tab:arti-fix} in Appendix \ref{sec:table}) has $m=50$ unique design points and there are $q=22$ effects.
According to Proposition \ref{prop:cond2}, the sufficient condition requires $n_0\geq 7$ and the necessary condition requires $n_0\geq 1$.
The local $D$-optimal design in Table \ref{tab:arti-fix} only satisfies the necessary condition.
To meet the sufficient condition $n$ has to be much larger.
For the global optimal design considering all possible $\bm \eta$ values, Proposition \ref{prop:cond2} can provide some guidelines for the design construction when $\bm\eta$ is varied in a relatively small range.

\section{Optimal Design under Conjugate Priors}\label{sec:conjugate}

When prior information for parameters $\bm \beta^{(1)}$, $\bm\beta^{(2)}$, and $\bm\eta$ is available, it would be desirable to consider the optimal design under the informative priors.
In this section, we detail the proposed Bayesian $D$-optimal design using the conjugate priors.

\subsection{Design Criterion}\label{subsec:d-criterion}
For the parameters $\bm \beta^{(1)}$ and $\bm \beta^{(2)}$, the conjugate priors are normal distribution since $Y|Z$ follows normal distribution.
Thus we consider their priors as
\[\bm \beta^{(1)}\sim N(\bm 0, \tau^2\bm R_1),\quad \bm \beta^{(2)}\sim N(\bm 0, \tau^2\bm R_2).\]
where $\tau^{2}$ is the prior variance and $\bm R_i$ is the prior correlation matrix of $\bm \beta^{(i)}$ for $i=1,2$.
Here we use the same prior variance $\tau^2$ only for simplicity.
The matrix $\bm R_i$ can be specified flexibly such as using $(\bm F'\bm F)^{-1}$, or those in \cite{joseph2006bayesian} for factorial designs.

For the parameters $\bm \eta$, we choose the conjugate prior derived in \cite{chen2003conjugate}.
It takes the form
\begin{equation}
\label{eq:eta-prior}
\bm \eta \sim D(s,\bm b) \propto \exp \left\{\sum_{i=1}^ns\left(b_i\bm f(\bm x_i)'\bm \eta-\log(1+e^{\bm f(\bm x_i)'\bm \eta})\right)\right\},
\end{equation}
where $D(s,\bm b)$ is the distribution with parameters $(s,\bm b)$.
Here $s$ is a scalar factor and $\bm b\in (0, 1)^n$ is the marginal mean of $\bm z$ as shown in \cite{diaconis1979conjugate}.
The value of $\bm b$ can be interpreted as a prior prediction (or guess) for $\mathbb{E}(\bm Z)$.
Based on the priors for $(\bm \beta^{(1)}, \bm \beta^{(2)}, \bm \eta)$ we can derive the posteriors as follows.
\begin{proposition}\label{prop:post}
For priors $\bm \beta^{(i)}\sim N(\bm 0, \tau^2\bm R_i)$\change{,} $i=1,2$ \change{,} and $\bm \eta\sim D(s, \bm b)$,
the posterior distributions of $\bm \beta^{(1)}$, $\bm \beta^{(2)}$ and $\bm \eta$ are independent of each other with the following forms.
\begin{align}
\bm \beta^{(i)}|\bm y, \bm z, \mu_0, \sigma^2, \bm X \sim  N &\left(\bm H_{i}^{-1}\bm F'\bm V_i(\bm y-\mu_0{\bf 1}), \sigma^2 \bm H_{i}^{-1} \right), \qquad \textrm{for } i=1, 2. \\
\bm \eta|\bm z, \bm X\sim D&\left(1+s, \frac{\bm z+s\bm b}{1+s}\right),
\end{align}
where $\bm H_{i} = \bm F'\bm V_i\bm F+ \rho \bm R_i^{-1}$ with $\rho = \frac{\sigma^2}{\tau^2}$.
\end{proposition}
The proof of Proposition \ref{prop:post} can be derived following the standard Bayesian framework, thus is omitted.
To derive the Bayesian $D$-optimal design criterion, we take the posterior distributions in Proposition \ref{prop:post} to \eqref{eq:gen_crit} and set $\psi(\cdot)=\log(\cdot)$.
The derivation is very similar to that in \eqref{eq:noninfo_Dopt_exact}, and thus we obtain the exact design criterion as
\begin{align}
\label{eq:conju_Dopt_exact}
\Psi(\bm X|\mu_0,\sigma^2)&=\mathbb{E}_{\bm z,\bm \eta}\left\{ \log(p(\bm\eta|\bm z, \bm X))\right\}\\\nonumber
&+\frac{1}{2}\sum_{i=1}^2\mathbb{E}_{\bm \eta}\mathbb{E}_{\bm z|\bm \eta}\left\{\log\det(\bm F'\bm V_i\bm F+\rho\bm R_i^{-1})\right\}+\textrm{constant}.
\end{align}
As the integration of $\mathbb{E}_{\bm z, \bm \eta}\left\{ \log(p(\bm\eta|\bm z, \bm X))\right\}$ is not tractable, we adopt the same normal approximation of the posterior distribution $p(\bm \eta|\bm z, \bm X)$ as in \eqref{eq:noninfo-approx1}.
A straightforward calculation leads to getting Fisher information matrix $\bm I(\bm \eta|\bm X)=(1+s)\bm F' \bm W_0 \bm F$.
Thus we have
\begin{equation}
\label{eq:conju-approx1}
\mathbb{E}_{\bm z,\bm \eta}\left\{ \log(p(\bm\eta|\bm z, \bm X))\right\}\approx \mathbb{E}_{\bm \eta}\{\log \det (\bm F'\bm W_0\bm F)\}+\textrm{constant}.
\end{equation}
Disregarding the constant, we can approximate the exact $\Psi(\bm X|\mu_0, \sigma^2)$ by
\[
\Psi(\bm X|\mu_0, \sigma^2)\approx \mathbb{E}_{\bm \eta}\{\log\det(\bm F'\bm W_0\bm F)\}+\frac{1}{2}\sum_{i=1}^2\mathbb{E}_{\bm \eta}\mathbb{E}_{\bm z|\bm \eta}\left\{\log\det(\bm F'\bm V_i\bm F+\rho\bm R_i^{-1})\right\},
\]
The following Theorem \ref{thm:info-d-criterion} gives an upper bound of the approximated criterion $\Psi(\bm X|\mu_0,\sigma^2)$ to avoid the integration with respect to $\bm z$, which plays the same role as Theorem \ref{the:noninfo-d-criterion}.
\begin{theorem}\label{thm:info-d-criterion}
Assume that the prior distributions of $\bm \beta^{(i)}$ are $\bm \beta^{(i)}\sim N(\bm 0, \tau^2\bm R_i)$ for $i=1,2$ and $\bm \eta$ has either the conjugate prior $\bm \eta\sim D(s, \bm b)$ or the noninformative prior $p(\bm \eta)\propto 1$.
If $\bm F'\bm W_0\bm F$  is nonsingular, an upper bound of the approximated $\Psi(\bm X|\mu_0, \sigma^2)$ is
\begin{equation}\label{eq:lower2}
Q(\bm X)=\mathbb{E}_{\bm\eta}\left(\log\det(\bm F'\bm W_0\bm F)+\frac{1}{2}\sum_{i=1}^2\log \det(\bm F'\bm W_i\bm F+\rho \bm R_i^{-1})\right).
\end{equation}
\end{theorem}
For the same argument as in Section \ref{sec:noninformative}, we use the upper bound in \eqref{eq:lower2} as the optimal design criterion.
\change{Note that since $\rho \bm R_i^{-1}$ is added to $\bm F'\bm V_i\bm F$ and $\bm F'\bm W_i\bm F$, $\bm F'\bm V_i\bm F+\rho\bm R_i^{-1}$ and $\bm F'\bm W_i\bm F+\rho\bm R_i^{-1}$ are nonsingular.}
The derivation of $\Psi(\bm X|\mu_0, \sigma^2)$ and $Q(\bm X)$ only needs $\bm F'\bm W_0\bm F$ to be nonsingular, which requires $m\geq q$ and $\bm \eta$ to be finitely bounded as in Section \ref{sec:noninformative}.

\subsection{Interpretation}\label{subsec:inter}

Note that the criterion in \eqref{eq:lower2} has a similar formulation with $Q(\bm X)$ in \eqref{eq:lower}.
The only difference is that \eqref{eq:lower} does not involve $\rho\bm R_i^{-1}$.
For consistency, we use the formula \eqref{eq:lower2} as the design criterion $Q(\bm X)$ for both cases.
When noninformative priors for $\bm\beta^{(1)}$ and $\bm \beta^{(2)}$ are used, we set $\rho=0$.
From another point of view, as $\tau^2\rightarrow \infty$, $\rho\rightarrow 0$, the variances in the priors $p(\bm \beta_1)$ and $p(\bm \beta_2)$ diffuse and result in a noninformative priors.

The criterion $Q(\bm X)$, consisting of three additive terms, can be interpreted intuitively.
The first additive term $\mathbb{E}_{\bm\eta}\{\log\det(\bm F'\bm W_0\bm F)\}$ is known as the Bayesian $D$-optimal criterion for logistic regression and $\mathbb{E}_{\bm\eta}\{\log\det(\bm F'\bm W_i\bm F+\rho\bm R_i^{-1})\}$ is the Bayesian $D$-optimal criterion for the linear regression model of $Y$.
To explain the weights, we rewrite $Q(\bm X)$ as follows.
\[
Q(\bm X)=1\cdot \mathbb{E}_{\bm\eta}\left(\log\det(\bm F'\bm W_0\bm F)\right)+1\cdot\left(\frac{1}{2}\sum_{i=1}^2\mathbb{E}_{\bm\eta}(\log \det(\bm F'\bm W_i\bm F+\rho \bm R_i^{-1}))\right)
\]
Since there are equal numbers of binary and continuous response observations, the design criterion should put the same weight (equal to 1) on both design criteria for $Z$ and $Y$.
For the two criteria for the linear regression models, the same weight 1/2 is used.
This is also reasonable because we assume $\pi_i\in (0, 1)$.
Then none of the diagonal entries of $\bm W_1$ and $\bm W_2$ are zero, so the two terms should split the total weight 1 assigned for the entire linear regression part.
Therefore, even though $Q(\bm X)$ are derived analytically, all the additive terms and their weights make sense intuitively.

\subsection{Prior Parameters}\label{subsec:prior_para}

Note that the conjugate prior $p(\bm \eta)$ requires prior parameters $(s, \bm b)$ to be specified.
Moreover, the prior distribution \eqref{eq:eta-prior} contains $f(\bm x_i)$, which depends on the design points.
When sampling $\bm \eta$ from the prior \eqref{eq:eta-prior}, it does not matter whether $f(\bm x_i)$'s are actually from the design points.
If relevant historical data is available, we can simply sample $\bm \eta$ from the likelihood of the data.
Alternatively, one can adopt the method in \cite{chen2003conjugate} to estimate the parameters $(s, \bm b)$.
Without the relevant data, we would use the noninformative prior for $\bm \eta$, i.e., $p(\bm \eta)\propto 1$ in the bounded region for $\bm \eta$.

The design criterion $Q(\bm X)$ contains some unknown parameters, including the noise-to-signal ratio $\rho=\sigma^2/\tau^2$ and the correlation matrices $\bm R_i$'s for $i=1,2$.
The value of $\rho$ has to be specified either from the historical data or from the domain knowledge. 
Typically we would assume $\rho<1$ such that the measurement error has a smaller variance than the signal variance.

The setting of $\bm R_i$ can also be specified flexibly.
If historical data are available, $\bm R_i$ can be set as the estimated correlation matrix of $\bm \beta^{(i)}$.
Otherwise, we can use the correlation matrix in \cite{joseph2006bayesian} and \cite{kang2009bayesian}, which is targeted for factorial design.
Specifically, let $\bm \beta$ be the unknown coefficients of the linear regression model and the prior distribution is $\bm \beta\sim N(\bm 0, \tau^2\bm R)$.
For 2-level factor coded in $-1$ and 1, \cite{joseph2006bayesian} suggests that $\bm R$ is a diagonal matrix and the priors for individual $\beta_j$ is
\begin{align}
\label{eq:prior-R}
&\beta_0\sim N(0, \tau^2), \\\nonumber
&\beta_j\sim N(0, \tau^2r),  \qquad i=1, \ldots, p,\\\nonumber
&\beta_j\sim N(0, \tau^2r^2), \qquad i= p+1, \ldots, p+\binom{p}{2},\\\nonumber
&\vdots\\\nonumber
&\beta_{2^p-1}\sim N(0, \tau^2r^p),
\end{align}
where $\beta_j$ $i=1, \ldots, p$ are main effects, $\beta_j$ $j=p+1,\ldots, p+\binom{p}{2}$ are 2-factor-interactions and up to the $p$-factor-interaction $\beta_{2^p-1}$.
The variance of $\beta_j$ decreases exponentially with the order of their corresponding effects by $r\in (0, 1)$, thus it incorporates the \emph{effects hierarchy principle} \citep{wu2011experiments}.
 \cite{joseph2006bayesian} showed that if $\bm f(\bm x)$ contains all the $2^p$ effects of all $p$ orders, $\tau^2\bm R$ can be represented alternatively by Kronecker product as  
$\tau^2\bm R=\varsigma^2\bigotimes_{j=1}^p\bm F_j(x_j)^{-1}\bm \Psi_j(x_j)(\bm F_j(x_j))^{-1}$.
The model matrix for the 2-level factor and the correlation matrix are
\begin{equation}
\label{eq:code2}
{\bm F_j(x_j)}=
\left(
\begin{array}{cc}
1 & -1 \\
1 & 1
\end{array}
\right) \;\; \textrm{and} \;\;\bm \Psi_j(x_j)=\left(
\begin{array}{cc}
1 & \zeta\\
\zeta & 1
\end{array}
\right).
\end{equation}
To keep the two different presentations equivalent, let $\zeta=\frac{1-r}{1+r}$ and $\tau^2=(\frac{1+\zeta}{2})^p\varsigma^2$.
For the mixed-level of 2- and 3-level experiments, \cite{kang2009bayesian} have extended the 2-level case to the format
\begin{equation}
\label{eq:prior-R2}
\tau^2\bm R=\varsigma^2\bigotimes_{j=1}^{p_2+p_{3,c}+p_{3,q}}\bm F_j(x_j)^{-1}\bm \Psi_j(x_j)(\bm F_j(x_j)^{-1})',
\end{equation}
where $p_2$ is the number of 2-level factors, $p_{3,c}$ is the number of 3-level qualitative (categorical) factors, and $p_{3,q}$ is the number of 3-level quantitative factors.
For all the 3-level factors, the model matrix is
\begin{equation}
\label{eq:code3}
{\bm F_j(x_j)}=
\left(
\begin{array}{ccc}
1 & -\sqrt{\frac{3}{2}} & \sqrt{\frac{1}{2}}\\
1 & 0 & -\sqrt{2}\\
1 & \sqrt{\frac{3}{2}} & \sqrt{\frac{1}{2}}
\end{array}
\right).
\end{equation}
An isotropic correlation function is recommended for the 3-level qualitative factors and a Gaussian correlation function for quantitative factors.
Thus, the correlation matrices for the 3-level qualitative and quantitative factors are
\begin{equation}
\label{eq:code3-corr}
{\bm \Psi_j(x_j)}=
\left(
\begin{array}{ccc}
1 & \zeta & \zeta\\
\zeta & 1 & \zeta\\
\zeta & \zeta & 1
\end{array}
\right) \;\; \textrm{and} \;\; \bm \Psi_j(x_j)=\left(
\begin{array}{ccc}
1 & \zeta & \zeta^4\\
\zeta & 1 & \zeta\\
\zeta^4 & \zeta & 1
\end{array}
\right),
\end{equation}
respectively.
To keep the covariance \eqref{eq:prior-R2} consistent with the 2-level case we still set $\zeta=\frac{1-r}{1+r}$.
To keep the variance of the intercept equal to $\tau^2$ \citep{kang2009bayesian}, we set
\[
\tau^2=\varsigma^2\left(\frac{1+\zeta}{2}\right)^{p_2}\left(\frac{1+2\zeta}{3}\right)^{p_{3,c}}\left(\frac{3+4\zeta+2\zeta^4}{9}\right)^{p_{3,q}},
\]
and thus
\[
\bm R=\left(\left(\frac{1+\zeta}{2}\right)^{p_2}\left(\frac{1+2\zeta}{3}\right)^{p_{3,c}}\left(\frac{3+4\zeta+2\zeta^4}{9}\right)^{p_{3,q}}\right)^{-1} \times \bigotimes_{j=1}^{p_2+p_{3,c}+p_{3,q}}\bm F_j(x_j)^{-1}\bm \Psi_j(x_j)(\bm F_j(x_j)^{-1})'.
\]
It is straightforward to prove that $\bm R$ is a diagonal matrix if only 2-level and 3-level qualitative factors are involved, but not so if any 3-level quantitative factors are involved, and the first diagonal entry of $\bm R$ is always 1.

To specify different prior distributions for $\bm \beta^{(1)}$ and $\bm \beta^{(2)}$, we only need to use different values $r_1$ (or $\zeta_1$) and $r_2$ (or $\zeta_2$) to construct the prior correlation matrix.
If the prior knowledge assumes that the two responses $Z$ and $Y$ are independent, one can set $r_1=r_2=r$ so that the two correlation matrices $\bm R_i$'s are the same, denoted as $\bm R$.
\cite{kang2009bayesian} has used $r=1/3$ (equivalently $\zeta=1/2$) according to a meta-analysis of 113 data sets from published experiments \citep{li2006regularities}.
Thus we also use $r=1/3$ in all the examples.
The readers can specify different values for $r_1$ and $r_2$ if needed.

In computation, we construct $\bm R$ using the Kronecker product in \eqref{eq:prior-R2}.
But such $\bm R$ is for $\bm f(\bm x)$ containing effects of all possible orders.
Usually, we would assume the model just contains lower-order effects.
So we just pick the rows and columns that correspond to the lower-order effects assumed in the model as the correlation matrix.

\section{Design Search Algorithm}\label{sec:alg}

In this work, we focus on the construction of optimal design based on factorial design, which is suited for the prior distribution introduced in Section \ref{subsec:prior_para}.
For optimizing the design criterion $Q(\bm X)$ we consider two cases.
First, for fixed $\bm \eta$ value, we develop a point-exchange algorithm to construct a \emph{local optimal design} that maximizes the criterion $Q(\bm X|\bm \eta)$.
Second, we construct a \emph{global optimal design} based on the prior distribution of $\bm \eta$.
Specifically, we construct the local optimal designs for different $\bm \eta$'s sampled from its prior distribution. Then the global optimal continuous design is obtained by accumulating the frequencies of design points selected into those local optimal designs.

\subsection{Local Optimal Design for Fixed $\eta$}\label{subsec:local}
For a fixed $\bm \eta$, we adapt the point-wise exchange algorithm to maximize the criterion
\[
Q(\bm X|\bm \eta)=\log \det(\bm F'\bm W_0\bm F)+\frac{1}{2}\sum_{i=1}^n \log\det(\bm F'\bm W_i\bm F+\rho\bm R_{i}^{-1}).
\]
The point-wise exchange algorithm is commonly used to construct $D$-optimal designs.
It was first introduced by \cite{fedorov1972theory} and then widely used in many works  \citep{cook1980comparison, nguyen1992review}.

The point-wise exchange algorithm finds the optimal design from a candidate set.
Here the candidate set is chosen to be the full factorial design without replicates.
For now, we develop the method for 2- and 3-level factors, but it can be generalized to factors of more levels.
Use previous notation that $p_2$, $p_{3, c}$, $p_{3,q}$ as the number of 2-level, 3-level categorical, and 3-level quantitative factors.
The total number of full factorial design points is $N=2^{p_2}3^{p_{3,c}+p_{3,q}}$, which can be large if the experiment involves many factors.
To make the algorithm efficient, we filter out the candidate points that are unlikely to be the optimal design points. Following the suggestion from \cite{dror2005approximate}, we exclude the candidate design points whose corresponding probabilities $\pi(\bm x, \bm \eta)$ is outside of $[0.15, 0.85]$. This range is used because the approximate variance of $\log\left(\frac{\pi_i}{1-\pi_i}\right)$ is nearly constant for $\pi_i \in (0.2, 0.8)$ but increases rapidly if $\pi_i$ is outside that range \citep{wu1985efficient}.
Denote the reduced candidate set as $\bm X_{c}$ with size $N'$.

Next we construct the initial design of size $n$, such that $\bm F'\bm W_0\bm F$ is nonsingular,
and so should be $\bm F'\bm W_i\bm F$ if $\rho=0$ for $i =1, 2$.
If $N'\geq q$, we construct the initial design by reduction.
Starting the initial design as $\bm X_{c}$, we remove the design points one by one until there are $q$ points left.
The remaining $n-q$ design points are then sampled from these $q$ initial design points with probabilities proportional to the lower bounds in the sufficient condition in Proposition \ref{prop:cond1}.
For removing one design point, we select the one having the smallest deletion function $d(\bm x)$ defined in \eqref{eq:deletion}.
Shortcut formulas are developed in Appendix for updating the inverse of the matrices $\bm F'\bm W_0\bm F$ and $\bm F'\bm W_i\bm F+\rho \bm R$ for $i=1, 2$ after one design point is removed.
If $N'\leq q$, we have to restore the candidate set back to the full factorial design and construct the initial design in the same reduction fashion.

To simplify the notation for $d(\bm x)$, we define $v_i(\bm x_1, \bm x_2)=\bm f(\bm x_1)'\bm M_i\bm f(\bm x_2)$ and $v_i(\bm x)=\bm f(\bm x)'\bm M_i\bm f(\bm x)$ for $i=0, 1, 2$, where $\bm M_0=\left(\bm F'\bm W_0\bm F\right)^{-1}$ and $\bm M_i=\left(\bm F'\bm W_i\bm F+\rho\bm R_i^{-1}\right)^{-1}$ for $i=1,2$.
Denote $\bm X$ as the current design and $\bm X_{-i}$ the design of $\bm X$ with the $i$th row removed.
Then the deletion function can be derived as
\begin{align}
\label{eq:deletion}
d(\bm x_i)&=Q(\bm X|\bm \eta)-Q(\bm X_{-i}|\bm \eta)\\\nonumber
&=-\left\{\log\left[1-\pi(\bm x_i, \bm \eta)(1-\pi(\bm x_i, \bm \eta))v_0(\bm x_i) \right]+\frac{1}{2}\log\left[1-\pi(\bm x_i, \bm \eta)v_1(\bm x_i)\right]\right.\\\nonumber
&\left.+\frac{1}{2}\log\left[1-(1-\pi(\bm x_i, \bm \eta))v_2(\bm x_i)\right]\right\}.
\end{align}
The smaller $d(\bm x_i)$ is, the less contribution the corresponding point makes for the overall objective $Q(\bm X|\bm \eta)$.

One key of the point-wise exchange algorithm is to compute $\Delta(\bm x, \bm x_i)=Q(\bm X^*|\bm \eta)-Q(\bm X|\bm \eta)$, the change in the criterion after the candidate design point $\bm x$ replaces $\bm x_i$ in the current design $\bm X$. Here $\bm X^*$ is the new design matrix after the exchange.
To compute $\Delta(\bm x, \bm x_i)$ efficiently, we can obtain the following formula.
\begin{align}
\label{eq:delta}
\Delta(\bm x, \bm x_i)&=Q(\bm X^{*}|\bm \eta)-Q(\bm X|\bm \eta)\\\nonumber
&=\log\Delta_0(\bm x, \bm x_i)+\frac{1}{2}\sum_{i=1}^2\log \Delta_i(\bm x, \bm x_i),
\end{align}
where $\Delta_i(\bm x, \bm x_i)$ for $i=0, 1, 2$ are derived in Appendix.
The matrices $\bm M_i$ for $i=0, 1, 2$ need to be updated after the exchange of design points.
Denote the updated matrices as $\bm M_i^*$ for the updated design $\bm X^*$.
We derive the shortcut formulas to easily compute $\bm M_i^*$ for $i=0, 1, 2$ as shown in Appendix.

Given the initial design, we can iteratively exchange the current design points with candidate design points to improve the objective $Q(\bm X|\bm \eta)$.
The details are listed in the following Algorithm \ref{alg:local}.
\begin{algorithm}
\caption{Exchange-Point Algorithm for Local $D$-Optimal Design}\label{alg:local}
\noindent

{\bf Step 0}: Generate the candidate design set from full factorial design.
Filter out the points with probabilities $\pi(\bm x, \bm \eta)$ outside of $[0.15, 0.85]$.

{\bf Step 1}: Generate the initial design.
Based on the initial design $\bm X$, update the matrices $\bm F$, $\bm W_i$, and $\bm M_i$ for $i=0, 1, 2$.
Compute the current objective value $Q(\bm X|\bm \eta)$.

{\bf Step 2}: Compute the deletion function $d(\bm x_i)$ for each $\bm x_{i}$ in $\bm X$.
Randomly sample one design point with probability inversely proportional to $d(\bm x_i)$'s.
Denote it as $\bm x_{i_0}$.

{\bf Step 3}: 
Find $\bm x^*$ as the candidate point having the largest $\Delta(\bm x,\bm x_{i_0})$.
If $\Delta(\bm x^*,\bm x_{i_0})>0$, exchange $\bm x^*$ with $\bm x_{i_0}$ in $\bm X$ and update the objective function value to $Q(\bm X|\bm \eta)+\Delta(\bm x^*, \bm x_{i_0})$.

{\bf Step 4}: Repeat \emph{Step 2} and \emph{3} until the objective function has been stabilized or the maximum number of iterations is reached.

\end{algorithm}

%
%
%
%

The Algorithm \ref{alg:local} can return different optimal designs due to different initial designs and the random sampling in Step 2.
Thus, we run Algorithm \ref{alg:local} a few times and return the design with the best optimal value. 
We have several remarks regarding the algorithm.
(1) The initial design generated via reduction does not have singularity issues.
(2) The updated design from point-exchange does not have the singularity problem either, based on the way $\bm x^*$ is selected and $\bm M_i^*$ for $i=0, 1, 2$ are computed.
(3) To avoid being trapped in a local maximum, in Step 2 we randomly sample the design point for an exchange instead of deterministically picking the ``worst'' point.
(4) Different from some other point-exchange algorithms, the candidate set here remains the same through Steps 1-4 since no points are deleted if they are selected in the design.
It enables the resultant optimal design having replicated design points.

\subsection{Global Optimal Design}\label{subsec:global}

Based on Algorithm \ref{alg:local} for local $D$-optimal design, we can use the following Algorithm \ref{alg:global} to construct global optimal design.
\begin{algorithm}
\caption{Algorithm for Global $D$-Optimal Design}\label{alg:global}
\noindent

{\bf Step 0}: If $p(\bm \eta)$ is informative, simulate $\bm \eta_j\sim p(\bm \eta)$ for $j=1,\ldots, B$.
Otherwise, $\bm \eta$ is uniformly distributed in a rectangular high-dimensional space.

{\bf Step 1}: For each $\bm \eta_j$, call \emph{Algorithm \ref{alg:local}} to construct the local optimal design $\bm X_j$.

{\bf Step 2}: For each point in the candidate set, count its frequency of being selected in the local optimal designs.
The continuous optimal design is formed by the normalized frequency as a discrete distribution.

{\bf Step 3}: To obtain a discrete optimal design, sample $n$ design points from the continuous optimal design.
\end{algorithm}

In Step 1 of generating $\bm \eta$ uniformly,
we can use uniform design \citep{fang2000uniform}, maximin Latin hypercube design \citep{morris1995exploratory}, or other space filling design methods \citep{joseph2008orthogonal,lin2010new,qian2012sliced} to select samples $\bm \eta_j$ for $j=1,\ldots, B$.
From Algorithm \ref{alg:global}, it is likely that the discrete design obtained in Step 3 has some design points with $n_i=1$.
When experimenters prefer to have replications at every design point, they can choose a saturated design by sampling $m=q$ unique design points in Step 3.
Then sample some $\bm \eta$ values as in Step 0.
Compute the lower bounds for $n_i$ for every $\bm \eta$ sample according to Proposition \ref{prop:cond1} and use the averaged lower bounds to set $n_i$.
If $\sum_{i=1}^mn_i$ exceeds $n$, the experimenters have to either increase the experiment budget or reduce the $\kappa$ value.

\section{Examples}\label{sec:egs}

In this section, we use two examples to demonstrate the proposed Bayesian $D$-optimal design and the construction method.
For both examples, we set $r=1/3$ (equivalently $\zeta=1/2$) as explained in Section \ref{subsec:prior_para}.
Since there are few existing works on experimental design for continuous and binary responses, we compare the proposed method with three alternative designs:
the optimal designs for the quantitative-only response, the optimal design for the binary-only response, and the naively combined design method as mentioned in Example 1.

\subsection{Artificial Example}\label{subsec:toy}

In this artificial experiment, there are three 2-level factors $x_1\sim x_3$, one 3-level categorical factor $x_4$, and one 3-level quantitative factor $x_5$.
The underlying model assumed is the complete quadratic model and $\bm f(\bm x)$ contains $q=22$ model effects including the intercept and the following model effects.
\begin{align*}
\textrm{First order effects:  }& x_1, x_2, x_3, x_{4,1}, x_{4, 2}, x_{5,l}, \\
\textrm{Second order effects:  } & x_1x_2, x_1x_3, x_1x_{4,1}, x_1x_{4, 2}, x_1x_{5,l}, x_2x_3, x_2x_{4,1}, x_2x_{4, 2}, \\
& x_2x_{5,l}, x_3x_{4,1}, x_3x_{4, 2}, x_3x_{5,l}, x_{4,1}x_{5,l}, x_{4, 2}x_{5,l}, \change{x_{5,\text{quad}}}.
\end{align*}
Here for the 3-level factors $x_{4}$ and $x_{5}$, the effects $x_{4,1}$ (1st comparison) and $x_{5, l}$ (linear effect) have values $\left\{-\sqrt{\frac{3}{2}}, 0, \sqrt{\frac{3}{2}}\right\}$,
and $x_{4,2}$ (2nd comparison) and \change{$x_{5, \text{quad}}$} (quadratic effect) have values $\left\{-\sqrt{\frac{1}{2}}, \sqrt{2}, \sqrt{\frac{1}{2}}\right\}$.
For the 2-level factors, the effects $x_i$ $i=1, 2, 3$ have the same values as the design settings $\{-1,1\}$.
We consider independent uniform distribution for each $\eta_i$. Specifically, $\eta_i\sim \textrm{Uniform}[-1, 1]$ for the intercept and the first order effects and Uniform$[-0.5,0.5]$ for the second order effects.
The ranges of $\eta_i$'s satisfy the effect hierarchy principle.
\begin{table}[h]
\begin{center}
\caption{An example of $\bm\eta$ value for the local $D$-optimal design.\label{tab:eta-local}}
\begin{tabular}{|c|c|c|c|c|c|c|c|}\hline
Effect & $\eta$ & Effect & $\eta$ &Effect & $\eta$ & Effect & $\eta$ \\\hline
Intercept & -0.0153 & $x_1$ & -0.6067  & $x_2$ & 0.7212 & $x_1x_2$ &  0.0080 \\\hline
$x_3$ & -0.1682  & $x_1x_3$ & 0.0010  & $x_2x_3$ & 0.1349 & $x_{4,1}$ & 0.0283 \\\hline
$x_1x_{4,1}$ & 0.0594 & $x_2x_{4,1}$ & -0.1719  & $x_3x_{4,1}$ & 0.1492 & $x_{4,2}$ & -0.1468\\\hline
$x_1x_{4,2}$ & 0.0553 & $x_2x_{4,2}$ & -0.0634 & $x_3x_{4,2}$ & -0.2629  & $x_{5,l}$ & -0.0660\\\hline
$x_1x_{5,l}$ & -0.1054 & $x_2x_{5,l}$ & -0.0857 & $x_3x_{5,l}$ & -0.0807 & $x_{4,1}x_{5,l}$ & -0.1198\\\hline
$x_{4,2}x_{5,l}$ & -0.0292 & $x_{5,\text{quad}}$ &  -0.1336 & \multicolumn{4}{c|}{}\\\hline
\end{tabular}
\end{center}
\end{table}

We set the experimental run size to be $n=66$. Table \ref{tab:eta-local} illustrates the values of a randomly chosen $\bm \eta$.
Using Algorithm \ref{alg:global} with this $\bm \eta$, we construct the proposed local $D$-optimal designs for QQ model $D_{QQ}$ for $\rho=0$ and $\rho=0.3$, respectively.
For comparison, we also generate three alternative designs via R package \emph{AlgDesign} developed by \cite{AlgDesign}.
Specifically, they are (i) the 66-run classic $D$-optimal design for linear regression model, denoted as $D_L$,
(ii) the 66-run local $D$-optimal design for logistic regression model given the $\bm \eta$, denoted as $D_G$,
(iii) and the naively combined design of $44$-run local $D$-optimal design for logistic regression model and 22-run $D$-optimal design for the linear regression model, denoted as $D_C$.
The details of these designs can be found in Table \ref{tab:arti-fix} in Appendix.

To evaluate the performance of the proposed design in comparison with alternative designs,
we consider the efficiency between two designs \citep{woods2006designs} as
\begin{align}\label{eq: efficiency}
\textrm{eff}(D_{1},D_{2}|\bm\eta)=\exp\left\{\frac{1}{q}\left(Q(D_{1}|\bm \eta)-Q(D_{2}|\bm \eta)\right)\right\}.
\end{align}
Table \ref{tab:arti-fix-opt} reports the efficiency of $D_{QQ}$ compared with $D_L$, $D_G$, and $D_C$, respectively.
The proposed QQ optimal design $D_{QQ}$ gains the best efficiency over the three alternative designs.
It appears that the combined design $D_C$ has the second-best design efficiency.
\begin{table}[h]
\begin{center}
\caption{Design efficiency between the proposed local design $D_{QQ}$ and three alternative designs. \label{tab:arti-fix-opt}}
\begin{tabular}{|c|c|c|c|}\hline
$\rho$ & $\textrm{eff}(D_{QQ},D_{L}|\bm\eta)$ & $\textrm{eff}(D_{QQ},D_{G}|\bm\eta)$ & $\textrm{eff}(D_{QQ},D_{C}|\bm\eta)$ \\ \hline
0 & 1.08 & 1.11 & 1.05 \\
\hline
0.3 & 1.10 & 1.14 & 1.07\\
\hline
\end{tabular}
\end{center}
\end{table}

\begin{figure}
\begin{center}
\includegraphics[scale=0.52]{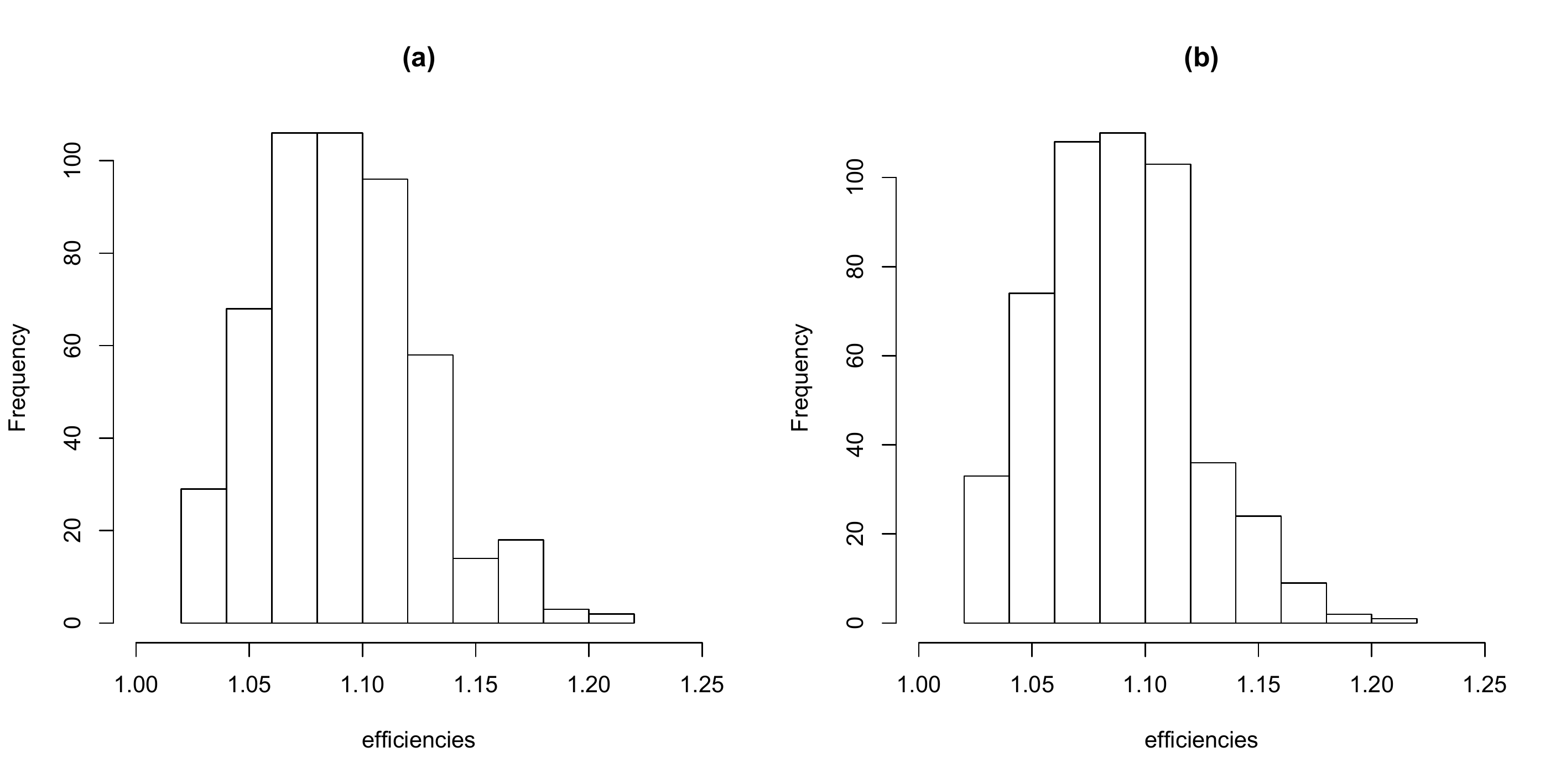}
\caption{Artificial example: the efficiency between each local $D_{QQ}$ and $D_C$ under different $\bm \eta$ values: (a) $\rho=0$ (b) $\rho=0.3$. \label{fig:arti-eff-local}}
\end{center}
\end{figure}

Next, we focus on the comparison of $D_{QQ}$ with $D_C$ under different $\bm \eta$ values.
We generate a maximin Latin hypercube design of $B=500$ runs (R package \emph{lhs} by \cite{lhs}) for $\bm \eta$ with the lower and upper bounds specified earlier.
For each of the $\bm\eta$ values, we construct a local QQ optimal design $D_{QQ}$ and the combined design $D_C$.
Figure \ref{fig:arti-eff-local} shows the histogram of the $\textrm{eff}(D_{QQ}, D_C|\bm \eta)$ for different $\bm \eta$ value.
All $\textrm{eff}(D_{QQ}, D_C|\bm \eta)$ values are larger than 1, indicating that the local QQ optimal design outperforms the combined design.

\begin{figure}
\begin{center}
\includegraphics[scale=0.65]{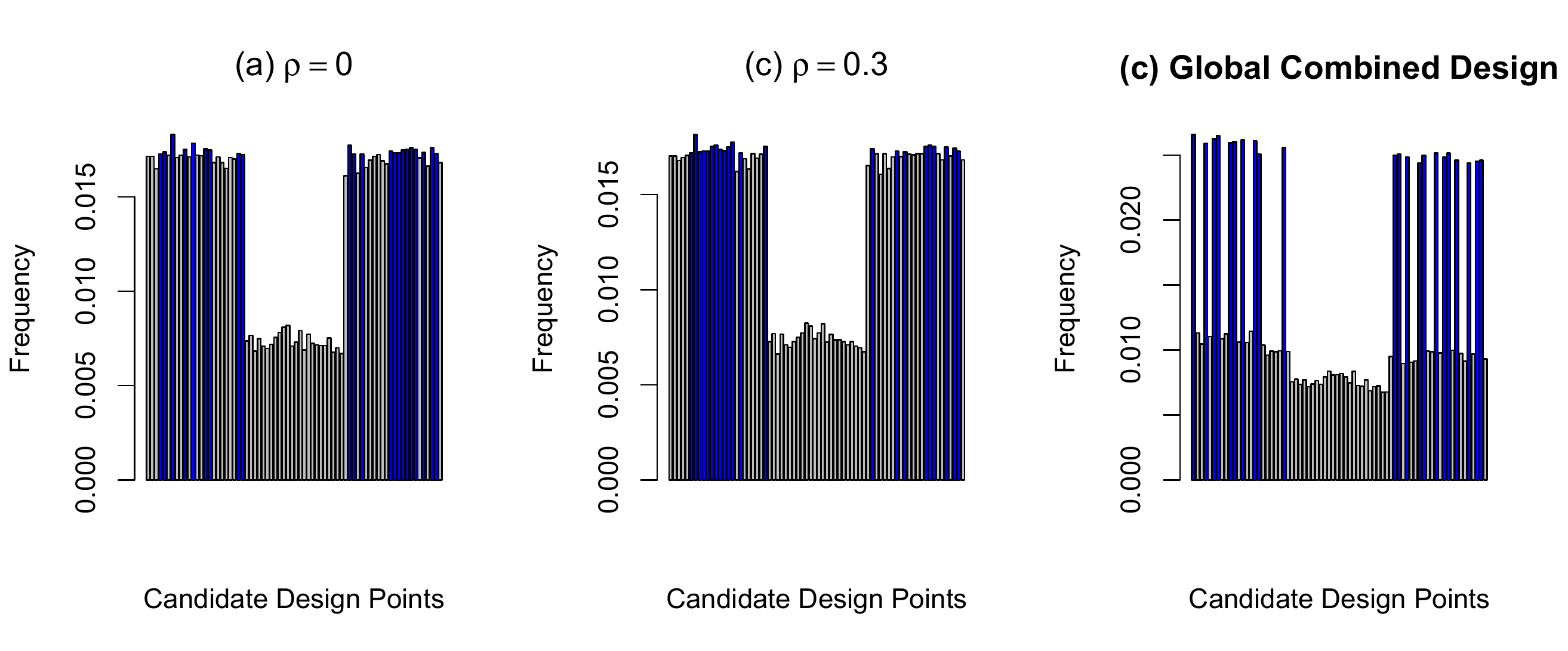}
\caption{Artificial example: global QQ optimal designs for (a) $\rho=0$ and (b) $\rho=0.3$ and (c) global combined design.\label{fig:arti-global}}
\end{center}
\end{figure}

Based on Algorithm \ref{alg:global}, we accumulate frequencies of locally optimal designs and obtain the global $D$-optimal designs shown in Figure \ref{fig:arti-global}.
Denote $d_{QQ}$ and $d_{C}$ are the proposed global optimal design for the QQ model and the global optimal combined design, respectively.
The bar plots show the normalized frequencies for all the candidate points with the largest 22 frequencies colored blue.
From Figure \ref{fig:arti-global}, for $d_{QQ}$, the points in the middle have much smaller frequencies than the other points.
It is known that these points in the middle correspond to the points with $x_5=0$ in Table \ref{tab:arti-fix}.
Note that these points are only necessary for estimating the coefficient for $x_{5,q}$, whose variances are the smallest in the prior for $\bm \beta^{(i)}$'s based on the effects hierarchy principle.
In contrast, such a pattern is not observed for points with $x_4=0$. 
The reason is that $x_4$ is a categorical variable and $x_4=-1, 0, 1$ are equally necessary to estimate the effects $x_{4,1}$ and $x_{4,2}$.
For $d_C$, the points with the largest 22 frequencies correspond to the 22-run $D$-optimal design for the linear regression model, which is independent of $\bm \eta$ and remains the same every time.
The points with $x_5=1$ and $-1$ only have slightly higher frequencies than the ones with $x_5=0$, due to the way we specify the prior distribution of $\bm \eta$.

\begin{figure}
\begin{center}
\includegraphics[scale=0.52]{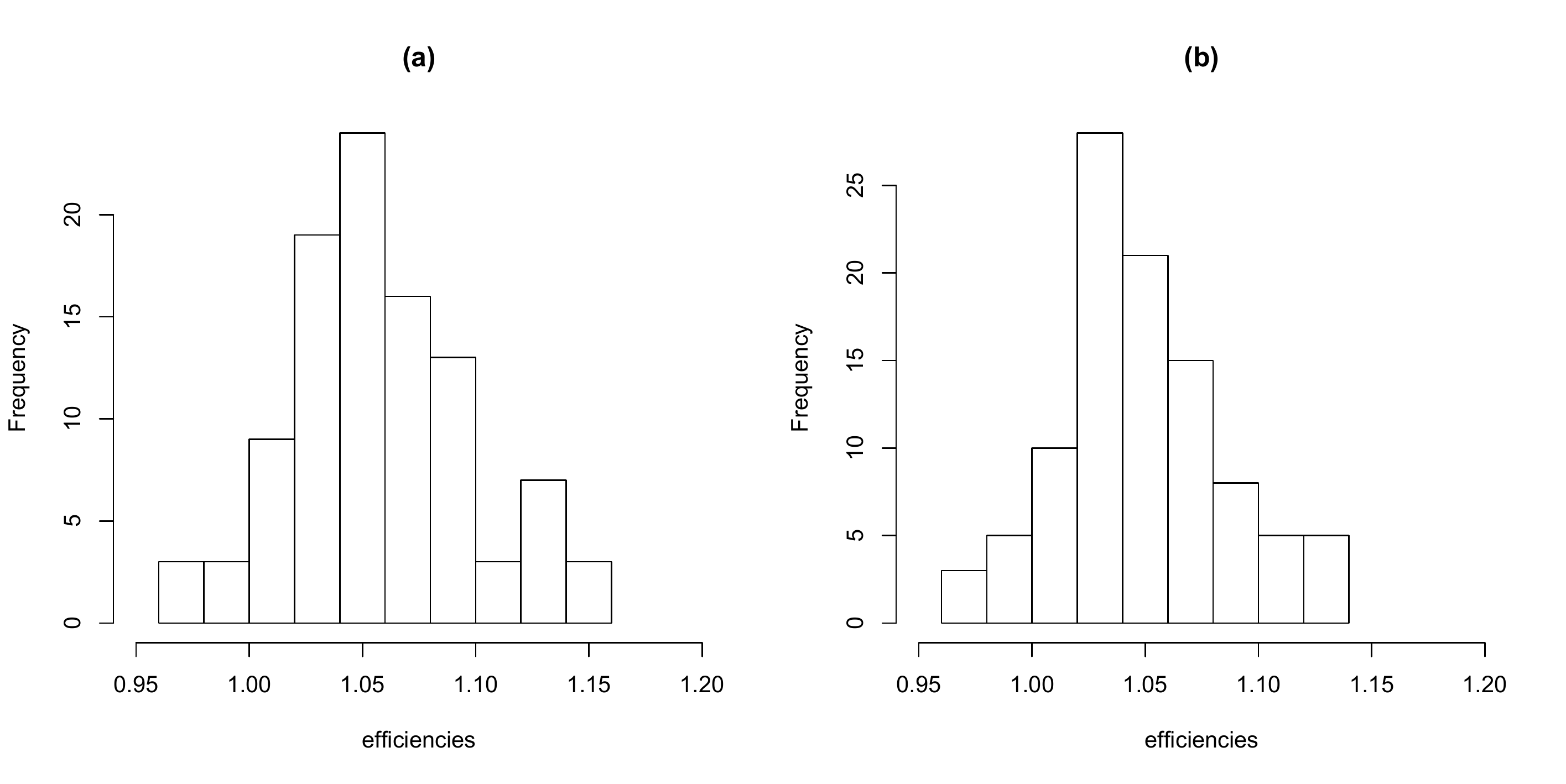}
\caption{Artificial example: the efficiency between each global QQ optimal design and combined design under different $\bm \eta$ values: (a) $\rho=0$ (b) $\rho=0.3$. \label{fig:arti-eff-global}}
\end{center}
\end{figure}

To compare the performances of the global designs, the design efficiencies in \eqref{eq: efficiency} is used with $Q(d|\bm \eta)$ adapted as 
\begin{align*}
Q(d|\bm \eta)&=\log\det\left(n\sum_{i=1}^N d(\bm x_i)\pi(\bm x_i,\bm\eta)(1-\pi(\bm x_i,\bm \eta))\bm f(\bm x_i)f(\bm x_i)'\right)\\
&+\frac{1}{2}\log\det\left(n\sum_{i=1}^N d(\bm x_i)\pi(\bm x_i, \bm \eta)\bm f(\bm x_i)\bm f(\bm x_i)'+\rho\bm R\right)\\
&+\frac{1}{2}\log\det\left(n\sum_{i=1}^N d(\bm x_i)(1-\pi(\bm x_i, \bm \eta))\bm f(\bm x_i)\bm f(\bm x_i)'+\rho\bm R\right)
\end{align*}
for a global optimal design $d$ given $\bm \eta$ value.
Here $d(\bm x_i)$ is the probability frequency for candidate design point $\bm x_i$ specified by the design $d$ and $\sum_{i=1}^N d(\bm x_i)=1$.
For the global optimal designs $d_{QQ}$ and $d_C$ obtained previously, Figure \ref{fig:arti-eff-global} shows the histograms of the $\textrm{eff}(d_{QQ}, d_C|\bm \eta)$ values, where the $\bm \eta$ values are generated from another 100-run maximin Latin hypercube design.
It is clear that $d_{QQ}$ is universally better than $d_C$, thus the proposed design is more robust to different values of $\bm \eta$.

\subsection{Etching Experiment}\label{subsec:etching}

In the etching process described in Section \ref{sec:intro}, 
the etchant is circulating at a certain flow rate. 
The wafers are rotated and swung horizontally and vertically.
Meanwhile, the air is blown in the etchant with certain pressure.
There are five factors involved in the etching process, the wafer rotation speed ($x_1$), the pressure for blowing the bubbles ($x_2$), the horizontal and vertical frequencies for swinging wafers ($x_3$, $x_4$), and the flow rate of circulating the etchant ($x_5$).
The engineers intend to experiment to study the relationship between these factors and the two QQ responses.

Because of the newly developed process, the historical data on similar processes are not directly applicable to this experiment.
Based on some exploratory analysis, we set $\rho=0.5$.
Both domain knowledge and data have shown that the wafer appearance is the worst when both the rotating speed ($x_1$) and bubble pressure ($x_2$) are low.
Accordingly, we set the prior of $\bm \eta$ as follows.
For intercept $\eta_0\sim\textrm{Uniform}[0,6]$. 
The linear effects of rotating speed and bubble pressure follow $\textrm{Uniform}[1, 5]$. 
The other linear effects follow $\textrm{Uniform}[-1, 1]$ and the second order interactions and quadratic effects $\textrm{Uniform}[-0.3, 0.3]$.
The experimental run size is set to be $n=21\times 6=126$, 6 times the number of effects $q=21$.

\begin{figure}
\begin{center}
\includegraphics[scale=0.52]{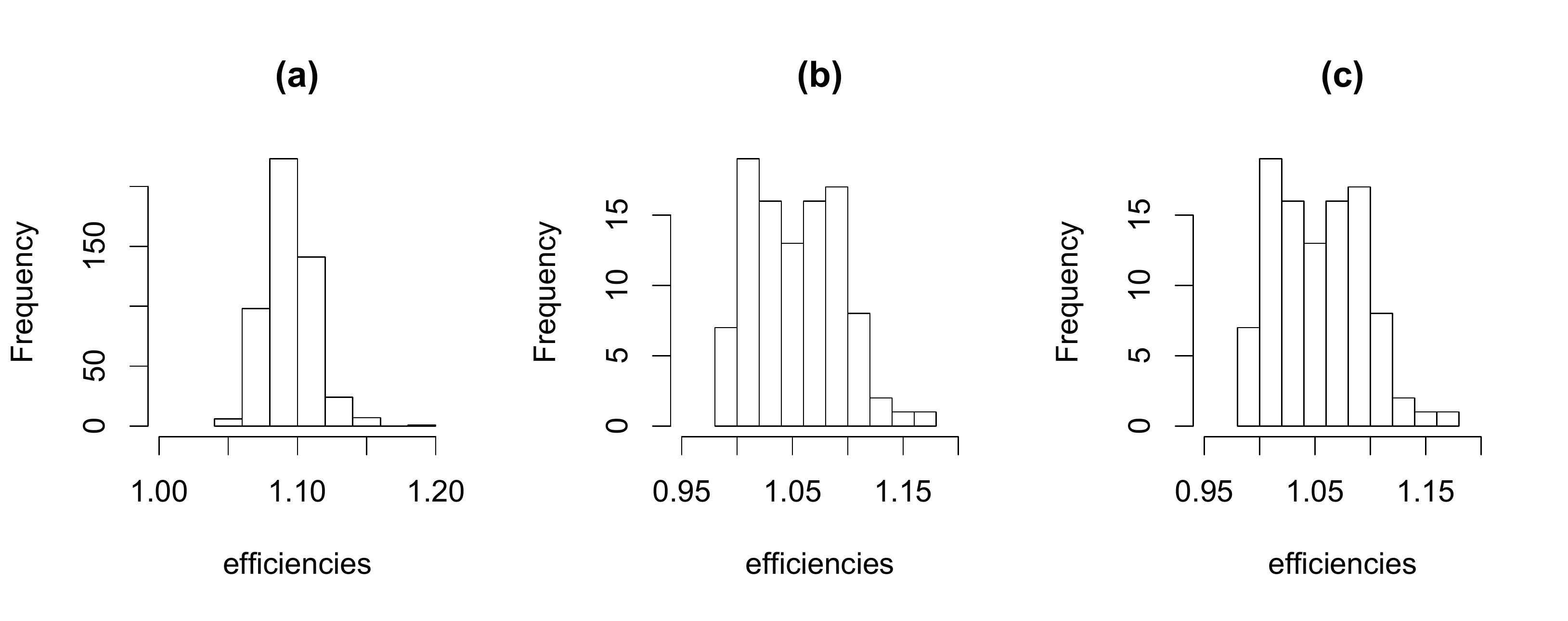}
\caption{Etching experiment: histograms of the efficiencies (a) efficiencies between local designs $D_{QQ}$ and $D_C$ for different $\bm \eta$'s; (b) efficiencies between global designs $d_{QQ}$ and $d_C$; (c) efficiencies between global design $d_{QQ}$ and the $3^{5-2}$ MA fractional factorial design. \label{fig:etch-eff}}
\end{center}
\end{figure}

We generate a maximin Latin hypercube design of $B=500$ runs for $\bm \eta$ values. For each $\bm \eta$ value, we obtain the local optimal designs $D_{QQ}$ and $D_C$.
Here the local combined design $D_C$ has $2/3$ of the runs generated from the local $D$-optimal design for logistic regression and $1/3$ of the runs from the $D$-optimal design for linear regression.
The efficiency between each pair of local designs $D_{QQ}$ and $D_C$ is reported in Figure \ref{fig:etch-eff}(a).
We can see that almost every local design $D_{QQ}$ is better than $D_C$.
Moreover, we obtain the global optimal designs $d_{QQ}$ and $d_C$ by accumulating the frequencies of the local designs.
To compare $d_{QQ}$ and $d_C$, we generate another 100-run maximin Latin hypercube design for $\bm \eta$ values and compute the efficiencies between $d_{QQ}$ and $d_C$ under different $\bm \eta$ values, which are shown in Figure \ref{fig:etch-eff} (b).
Clearly, $d_{QQ}$ is universally better and more robust to $\bm \eta$ than $d_C$.

Fractional factorial design \citep{wu2011experiments}  is another commonly used design in practice.
We compare the proposed design with a $3^{5-2}$ minimum aberration (MA) fractional factorial design by defining the contrast subgroup as $I=ABD^2=AB^2CE^2=AC^2DE=BCDE^2$.
Each design point is replicated 5 times and the overall run is $3^{5-2}\times 5=135$. Figure \ref{fig:etch-eff}(c) shows the histogram of the efficiencies between $d_{QQ}$ and the MA design, and the proposed global optimal design is still superior.

\begin{figure}
\begin{center}
\includegraphics[scale=0.6]{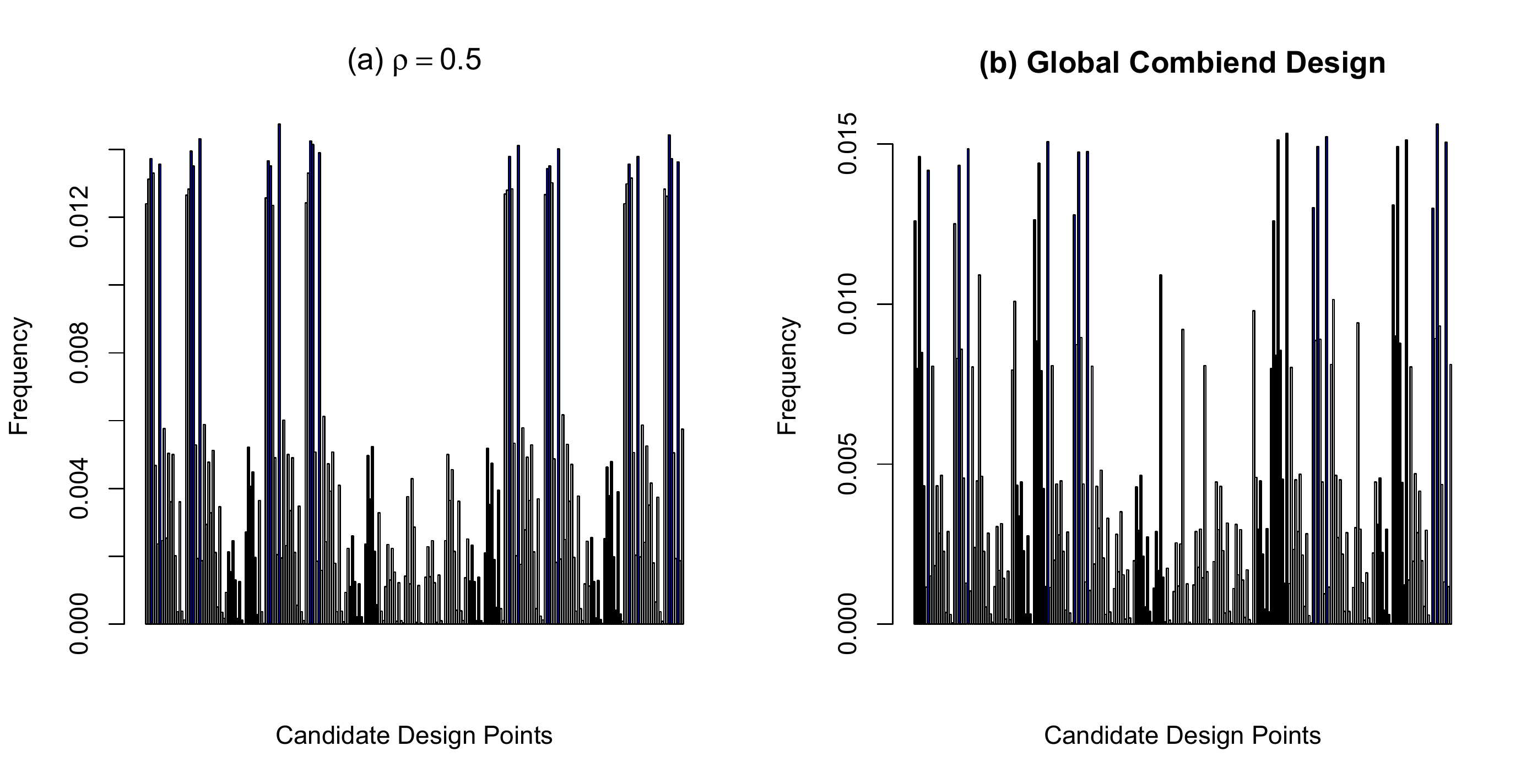}
\caption{Etching experiment: (a) the global Bayesian QQ $D$-optimal design for $\rho=0.5$ and (b) the global combined design.\label{fig:etch-doe}}
\end{center}
\end{figure}

\section{Discussion}\label{sec:end}

In this paper, we propose the Bayesian $D$-optimal design criterion for QQ responses.
The adoption of the Bayesian approach allows us to consider both the non-informative priors as the frequentist approach and informative priors when domain knowledge or historical data are available.
A new point-exchange algorithm is developed for efficiently constructing the proposed designs. This algorithm can also be used to construct non-factorial designs when the candidate set is not a full factorial design.
Moreover, the proposed method can be directly generalized for the sequential design with the QQ response.
In the following, we discuss some other scenarios for the proposed method that are not considered in detail previously.

\noindent{\bf Non-conjugate Prior for $\bm \eta$}

Other than the conjugate prior $p(\bm \eta)$, we can also use a non-conjugate prior distribution $\bm \eta\sim N(\bm 0, \tau_0^2 \bm R_0)$.
In this situation, one can consider the normal approximation in the posterior distribution $p(\bm \eta|\bm z)$.
Then the design criterion for the binary response becomes \citep{chaloner1995bayesian} $\mathbb{E}_{\bm z, \bm \eta}\{\log(p(\bm \eta|\bm z)\}\approx \mathbb{E}_{\bm \eta}\left\{\log \det(\bm F'\bm W_0\bm F+\rho_0\bm R_0^{-1})\right\}$.
The overall design criterion $Q(\bm X)$ can be updated as
\[
Q(\bm X)=\sum_{i=0}^2\mathbb{E}_{\bm \eta}\left\{\log\det\left(\bm F'\bm W_i\bm F+\tau_i\bm R_i^{-1}\right)\right\},
\]
where $\rho_i=\sigma^2/\tau_i$, $\tau_i$ and $\bm R_i$ are the prior variance and correlation of $\bm \eta$, $\bm \beta^{(1)}$, and $\bm \beta^{(2)}$ respectively.
The proposed design construction algorithm can still be applied with minor modifications.

\noindent{\bf Multiple QQ Responses}

In this paper, we focus on optimal designs for one quantitative response and one qualitative response.
The proposed method can also be generalized to accommodate the QQ models with multiple quantitative responses $Y_1,\ldots, Y_l$ and binary responses $Z_1,\ldots, Z_k$.
For example, a multi-level qualitative response can be transformed into a set of dummy binary responses.
One idea is to generalize the QQ models in \eqref{eq:model-Z} and \eqref{eq:model-Y} for both $l\geq 1$ and $k\geq 1$.
For $l\geq 1$ and $k=1$, we generalize the QQ model by introducing the correlation matrix between $Y_1,\ldots, Y_l$ as in the standard multi-response regression \citep{breiman1997predicting}.
Then the corresponding optimal design can be established by studying its likelihood function.
For $l\ge1$ and $k>1$ with multiple binary responses,
considering all $2^k$ conditional models for $(Y_1,\ldots, Y_l|Z_1=1, \ldots, Z_k=1)$,\ldots, $(Y_1,\ldots, Y_l|Z_1=0, \ldots, Z_k=0)$ only works for a small $k$.
Moreover, the construction algorithm can be more complicated as it needs to involve the multi-logit model \citep{mccullagh1980} for modeling the multiple binary responses.
When $k$ is relatively large, we are going to pursue an alternative QQ model and develop its corresponding optimal design method as a future research topic.

\noindent{\bf Continuous Design}

The point-exchange algorithm is to construct the exact discrete optimal designs, which are different from the theoretical continuous optimal designs.
As described in Sections 5 and 6, the way of generating the frequency as the local optimal design is heuristic.
The rigorous definition of the local continuous $D$-optimal design criterion is the probability measure $\xi$ on the design space $\Omega$ that maximizes
\begin{align*}
&Q(\bm X|\bm \eta)=\log\det\left(\int\pi(\bm x)(1-\pi(\bm x))f(\bm x)f(\bm x)'d\xi(\bm x)\right) + \\
&\log\det\left(\int(\pi\left(\bm x)f(\bm x)f(\bm x)'+\rho \bm R_1^{-1}\right)d\xi(\bm x)\right)+\log\det\left(\int\left((1-\pi(\bm x))f(\bm x)f(\bm x)'+\rho \bm R_2^{-1}\right)d\xi(\bm x)\right).
\end{align*}
\cite{yang2013optimal} developed a method to obtain the optimal $\xi$ for the nonlinear models.
It will be interesting to extend their framework and develop the method to obtain the optimal $\xi$ for QQ models.

\noindent {\bf Different QQ Models}

The proposed design is not restricted to the logit model for the binary response. 
For example, if the probit model is used,
the Bayesian $D$-optimal design criterion can be directly obtained by replacing the logit transformation with the probit transformation in both $p(\bm z|\bm \eta)$ and $p(\bm \eta)$.
The design criterion can be derived similarly with minor modifications.
The criterion formula remains the same with the following different diagonal matrices,
\begin{align*}
\bm W_0&=\diag\left\{\frac{\phi^2(\bm f(\bm x_1)'\bm \eta)}{\Phi(\bm f(\bm x_1)'\bm \eta)\left(1-\Phi(\bm f(\bm x_1)'\bm \eta)\right)}, \ldots, \frac{\phi^2(\bm f(\bm x_n)'\bm \eta)}{\Phi(\bm f(\bm x_n)'\bm \eta)\left(1-\Phi(\bm f(\bm x_n)'\bm \eta)\right)}\right\}, \\
\bm W_1&=\diag\left\{\Phi\left(\bm f(\bm x_1)'\bm \eta\right), \ldots, \Phi\left(\bm f(\bm x_n)'\bm \eta\right)\right\},\quad \bm W_2=\bm I-\bm W_1,
\end{align*}
where $\Phi$ and $\phi$ are CDF and PDF of the standard normal distribution.

It is worth pointing out that the design criterion in the work is based on the QQ model constructed by the joint model of $Y|Z$ in \eqref{eq:model-Z} and $Z$ in \eqref{eq:model-Y}.
\cite{kang2021auxillary} created a new QQ model based on $Z|U$ where $U$ is a latent continuous variable that is assumed to be correlated with the observed continuous response variable $Y$. 
Besides the conditional model structures, other model structures such as mixed graphical models \citep{yang2014mixed} can also be used as long as the $D$-optimality can be derived.

\begin{center}
{\bf \Large Acknowledgement}\\
\end{center}
The authors were partly supported by U.S. National Science Foundation for this research project. 
Dr. Lulu Kang was supported by grants CMMI-1435902, DMS-1916467, and DMS-2153029, Dr. Xinwei Deng by CMMI-1233571 and CMMI-1435996, and Dr. Ran Jin by CMMI-1435996. 

\bibliography{References}

\newpage

\begin{center}
{\Large \bf Appendix: Proofs, Derivations, and Extra Table}
\end{center}

\setcounter{figure}{0}
\setcounter{table}{0}
\setcounter{lem}{0}
\setcounter{theorem}{0}
\setcounter{proposition}{0}
\setcounter{section}{0}

\makeatletter 
\renewcommand{\thefigure}{A\@arabic\c@figure}
\renewcommand{\thetable}{A\@arabic\c@table}
\renewcommand{\thelem}{A\@arabic\c@lem}
\renewcommand{\theproposition}{A\@arabic\c@proposition}
\renewcommand{\thetheorem}{A\@arabic\c@theorem}
\renewcommand{\thesection}{A\@arabic\c@section}
\makeatother

\section{Proofs and Derivations}\label{sec:proofs}

\noindent{\bf Proof of \eqref{eq:noninfo_Dopt_exact}}
\begin{proof}
\begin{align*}
\Psi(\bm X|\mu_0,\sigma^2)=\int p(\bm y, \bm z|\mu_0,\sigma^2)&\int \log\left(p(\bm \beta^{(1)}, \bm \beta^{(2)}, \bm \eta|\bm y, \bm z,\mu_0, \sigma^2)\right)\\
&p(\bm \beta^{(1)}, \bm \beta^{(2)}, \bm \eta|\bm y, \bm z,\mu_0, \sigma^2)d\bm \beta^{(1)}d\bm\beta^{(2)}d\bm \eta d\bm yd\bm z\\
=\int p(\bm y, \bm z|\mu_0,\sigma^2)&\left\{\int \log\left(p(\bm \eta|\bm z) \right)p(\bm \eta|\bm z)d\bm \eta\right.\\
&+\left.\sum_{i=1}^2\int \log\left(p(\bm \beta^{(i)}|\bm y, \bm z, \mu_0,\sigma^2)\right)p(\bm \beta^{(i)}|\bm y, \bm z, \mu_0,\sigma^2) d\bm \beta^{(i)}\right\}d\bm y d\bm z\\
=\int p(\bm y, \bm z|\mu_0,\sigma^2)&\left\{\int \log\left(p(\bm \eta|\bm z) \right)p(\bm \eta|\bm z)d\bm \eta\right.\\
&-\left.\frac{1}{2}\sum_{i=1}^2\log \det\{\sigma^{-2}(\bm F'\bm V_i\bm F)^{-1}\}-n\log(2\pi)-n\right\}d\bm y d\bm z\\
=\int \log(p(\bm\eta|\bm z))p(\bm \eta|\bm z)&p(\bm z)d\bm \eta d\bm z\\
&-\frac{1}{2}\sum_{i=1}^2\int\log\det\{\sigma^{-2}(\bm F'\bm V_i\bm F)^{-1}\}p(\bm z)d\bm z+\textrm{constants} \\
=\int \log(p(\bm\eta|\bm z))p(\bm z, \bm \eta)&d\bm \eta d\bm z\\
&+\frac{1}{2}\sum_{i=1}^2\int\log\det\{(\bm F'\bm V_i\bm F)\}p(\bm z|\bm \eta)p(\bm \eta)d\bm z d\bm \eta+\textrm{constants}.
\end{align*}
Write the integration into the form of expectation,
\begin{align*}
\Psi(\bm X|\mu_0,\sigma^2)&=\mathbb{E}_{\bm z, \bm \eta}\left\{ \log(p(\bm\eta|\bm z))\right\}\\
&+\frac{1}{2}\sum_{i=1}^2\mathbb{E}_{\bm \eta}\mathbb{E}_{\bm z|\bm \eta}\left\{\log\det(\bm F'\bm V_i\bm F)\right\}+\textrm{constants}.
\end{align*}
\end{proof}

\noindent{\bf Proof of Theorem \ref{the:noninfo-d-criterion}}
\begin{proof}
To show \eqref{eq:lower}, we just need to show that $\mathbb{E}_{\bm z|\bm \eta}\left(\log \det(\bm F'\bm V_i\bm F)\right)\leq\log \det(\bm F'\bm W_i\bm F')$ for $i=1, 2$.

First we need to show that $\log\det(\bm F'\bm A\bm F)=\log\det(\sum_{i=1}^n a_i\bm f(\bm x_i)\bm f(\bm x_i)')$ with $\bm A=\diag\{a_1, \ldots, a_n\}$ is a concave function of $\bm a=(a_1, \ldots, a_n)'$ for $\bm a\in [0, 1]^n$.
Denote $\bm B=\diag\{b_1, \ldots, b_n\}$ for $\bm b=(b_1, \ldots, b_n)'\in [0, 1]^n$. We assume that $\bm F'\bm A\bm F$ is nonsingular, thus $\bm F'\bm A\bm F$ is positive definite.
For any scalar value of $t$, define function $g(t)$
\begin{align*}
g(t)&=\log\det\left(\bm F'\bm A\bm F+\bm F'(t\bm B)\bm F\right)\\
&=\log\det(\bm F'\bm A\bm F)+\log\det(\bm I_q+t(\bm F'\bm A\bm F)^{-1/2}(\bm F'\bm B\bm F)(\bm F'\bm A\bm F)^{-1/2})\\
&=\log\det(\bm F'\bm A\bm F)+\sum_{i=1}^q\log(1+t\lambda_i),
\end{align*}
where $\lambda_i$'s are the eigenvalues of the positive definite matrix $(\bm F'\bm A\bm F)^{-1/2}(\bm F'\bm B\bm F)(\bm F'\bm A\bm F)^{-1/2}$.
Thus $g(t)$ is a concave function in $t$ for any choice of $\bm a$, which is the sufficient and necessary condition that $\log\det (\bm F'\bm A\bm F)$ is a concave function of $\bm a$.
According to Jensen's inequality, if $\pi(\bm x_i, \bm \eta)\in (0, 1)$ for $i=1, \ldots, n$, then
\begin{align*}
\mathbb{E}_{\bm z|\bm \eta}\left(\log \det(\bm F'\bm V_1\bm F)\right)&\leq \log \det\left(\sum_{i=1}^n \mathbb{E}(Z_i|\bm \eta)\bm f(\bm x_i)\bm f(\bm x_i)'\right)\\
&=\log \det\left(\sum_{i=1}^n \pi(\bm x_i,\bm \eta)\bm f(\bm x_i)\bm f(\bm x_i)'\right), \\
\mathbb{E}_{\bm z|\bm \eta}\left(\log \det(\bm F'\bm V_2\bm F)\right)&\leq \log \det\left(\sum_{i=1}^n \mathbb{E}((1-Z_i)|\bm \eta)\bm f(\bm x_i)\bm f(\bm x_i)'\right)\\
&=\log \det\left(\sum_{i=1}^n(1- \pi(\bm x_i,\bm \eta))\bm f(\bm x_i)\bm f(\bm x_i)'\right).\\
\end{align*}
So we have proved \eqref{eq:lower}.
\end{proof}

\noindent{\bf Proof of Proposition \ref{prop:cond1}}
\begin{proof}
\[
m=\sum_{i=1}^m I\left(0<\sum_{j=1}^{n_i}Z_{ij}<n_i\right)+\sum_{i=1}^m I\left(\sum_{j=1}^{n_i}Z_{ij}=0\right)+\sum_{i=1}^m I\left(\sum_{j=1}^{n_i}Z_{ij}=n_i\right)\geq q
\]
If $m=q$, \eqref{eq:iffcond} is equivalent to $I\left(0<\sum_{j=1}^{n_i}Z_{ij}<n_i\right)=1$ for $i=1, 2, \ldots, m$.
A sufficient condition for any $\Pr\left(0<\sum_{j=1}^{n_i} Z_{ij}<n_i\right)\geq \kappa$ can be derived as follows.
\begin{align*}
\Pr\left(0<\sum_{j=1}^{n_i} Z_{ij}<n_i\right) &=1-\pi(\bm x_i, \bm \eta)^{n_i}-(1-\pi(\bm x_i, \bm \eta))^{n_i}\geq \kappa\\
\Longleftrightarrow \quad &\pi(\bm x_i, \bm \eta)^{n_i}+(1-\pi(\bm x_i, \bm \eta))^{n_i}\leq 1-\kappa.
\end{align*}
Next develop an upper bound for $\pi(\bm x_i, \bm \eta)^{n_i}+(1-\pi(\bm x_i, \bm \eta))^{n_i}$.
If $\pi(\bm x_i,\bm \eta)\geq 1/2$, denote $a=(1-\pi(\bm x_i, \bm \eta))/\pi(\bm x_i, \bm \eta)$ and $a\leq 1$.
Then
\[
\pi(\bm x_i, \bm \eta)^{n_i}+(1-\pi(\bm x_i, \bm \eta))^{n_i}=\frac{1+a^{n_i}}{(1+a)^{n_i}}\leq \frac{1+a}{(1+a)^{n_i}}.
\]
If
\[
\frac{1+a}{(1+a)^{n_i}}\leq 1-\kappa \quad \Longleftrightarrow \quad n_i\geq 1+\frac{\log(1-\kappa)}{\log \pi(\bm x_i, \bm \eta)},
\]
then $\pi(\bm x_i, \bm \eta)^{n_i}+(1-\pi(\bm x_i, \bm \eta))^{n_i}\leq 1-\kappa$.
If $\pi(\bm x_i,\bm \eta)< 1/2$, denote $a=\pi(\bm x_i, \bm \eta)/(1-\pi(\bm x_i, \bm \eta))$ and $a<1$. Then
\[
\pi(\bm x_i, \bm \eta)^{n_i}+(1-\pi(\bm x_i, \bm \eta))^{n_i}=\frac{1+a^{n_i}}{(1+a)^{n_i}}< \frac{1+a}{(1+a)^{n_i}}.
\]
The sufficient condition becomes
\[
\frac{1+a}{(1+a)^{n_i}}\leq 1-\kappa \quad \Longleftrightarrow \quad n_i\geq 1+\frac{\log(1-\kappa)}{\log (1-\pi(\bm x_i, \bm \eta))}.
\]
Combining the two cases, the sufficient condition on $n_i$ for $i=1, \ldots, m$ is \eqref{eq:regcond1_s}.
It is known that
\[2\left(\pi(\bm x_i,\bm \eta)(1-\pi(\bm x_i,\bm \eta)\right)^{n_i/2}\leq \pi(\bm x_i, \bm \eta)^{n_i}+(1-\pi(\bm x_i, \bm \eta))^{n_i}\leq 1-\kappa.\]
The necessary condition is
\[
2\left(\pi(\bm x_i,\bm \eta)(1-\pi(\bm x_i,\bm \eta)\right)^{n_i/2}\leq 1-\kappa \quad\Longleftrightarrow \quad n_i\geq \frac{2\log\left(\frac{1-\kappa}{2}\right)}{\log\pi(\bm x_i,\bm \eta)+\log(1-\pi(\bm x_i, \bm \eta))}.
\]
\end{proof}

\noindent{\bf Proof of Proposition \ref{prop:cond2}}
\begin{proof}
If $m>q$, \eqref{eq:iffcond} is equivalent to
\[
\sum_{i=1}^m I\left(\sum_{j=1}^{n_i}Z_{ij}=0\right)\leq m-q, \textrm{ and } \sum_{i=1}^m I\left(\sum_{j=1}^{n_i}Z_{ij}=n_i\right)\leq m-q.
\]
For the two inequalities to hold with large probability,
\begin{align}
\label{eq:regcond2_1}
& \sum_{i=1}^m \mathbb{E}\left\{I\left(\sum_{j=1}^{n_i}Z_{ij}=0\right)\right\}=\sum_{i=1}^m (1-\pi(\bm x_i,\bm \eta))^{n_i}\leq m-q,\\
\label{eq:regcond2_2}
\textrm{ and } &\sum_{i=1}^m \mathbb{E}\left\{I\left(\sum_{j=1}^{n_i}Z_{ij}=n_i\right)\right\}=\sum_{i=1}^m \pi(\bm x_i,\bm \eta)^{n_i}\leq m-q.
\end{align}
It is known that
\[
m\left(\prod_{i=1}^m (1-\pi_{\max})^{n_i}\right)^{1/m}\leq \sum_{i=1}^m(1-\pi_{\max})^{n_i}\leq \sum_{i=1}^m (1-\pi(\bm x_i,\bm \eta))^{n_i}\leq \sum_{i=1}^m (1-\pi_{\min})^{n_i}\leq m(1-\pi_{\min})^{n_0}.
\]
Thus one sufficient condition for \eqref{eq:regcond2_1} is
\[
m(1-\pi_{\min})^{n_0}\leq m-q \quad \Longleftrightarrow \quad n_0\geq \frac{\log(1-q/m)}{\log (1-\pi_{\min})}.
\]
One necessary condition for \eqref{eq:regcond2_1} is
\[m\left(\prod_{i=1}^m (1-\pi_{\max})^{n_i}\right)^{1/m}\leq m-q\quad \Longleftrightarrow \sum_{i=1}^m n_i\geq m\frac{\log(1-q/m)}{\log (1-\pi_{\max})}.\]
Similarly,
\[
m\left(\prod_{i=1}^m \pi_{\min}^{n_i}\right)^{1/m}\leq \sum_{i=1}^m \pi_{\min}^{n_i}\leq \sum_{i=1}^m \pi(\bm x_i,\bm \eta)^{n_i}\leq \sum_{i=1}^m\pi_{\max}^{n_i}\leq m\cdot\pi_{\max}^{n_0}.
\]
Thus one sufficient condition for \eqref{eq:regcond2_2} is
\[
 m\cdot\pi_{\max}^{n_0}\leq m-q \quad \Longleftrightarrow \quad n_0\geq \frac{\log(1-q/m)}{\log \pi_{\max}}.
\]
One necessary condition for \eqref{eq:regcond2_2} is
\[m\left(\prod_{i=1}^m \pi_{\min}^{n_i}\right)^{1/m}\leq m-q\quad \Longleftrightarrow \sum_{i=1}^m n_i\geq m\frac{\log(1-q/m)}{\log \pi_{\min}}.\]
Thus the sufficient condition for \eqref{eq:regcond2_1} and \eqref{eq:regcond2_2} derived here is
\[
n_0\geq \max\left\{1, \frac{\log(1-q/m)}{\log(1-\pi_{\min})}, \frac{\log(1-q/m)}{\log\pi_{\max}}\right\},
\]
or equivalently,
\[
\sum_{i=1}^m n_i\geq m\cdot n_0\geq m\cdot\max\left\{1, \frac{\log(1-q/m)}{\log(1-\pi_{\min})}, \frac{\log(1-q/m)}{\log\pi_{\max}}\right\}.
\]
The necessary condition for \eqref{eq:regcond2_1} and \eqref{eq:regcond2_2} derived here is
\[
\sum_{i=1}^n n_i\geq m \cdot \max\left\{1, \frac{\log(1-q/m)}{\log(1-\pi_{\max})}, \frac{\log(1-q/m)}{\log\pi_{\min}}\right\}.
\]
It is easy to see that this lower bound in the necessary condition is smaller than the one in the sufficient condition,
\[
\sum_{i=1}^m n_i\geq m\cdot n_0\geq m\cdot \max\left\{1, \frac{\log(1-q/m)}{\log(1-\pi_{\min})}, \frac{\log(1-q/m)}{\log\pi_{\max}}\right\}.
\]
\end{proof}

\noindent{\bf Proof of Theorem \ref{thm:info-d-criterion}}
\begin{proof}
We only need to show that $\log\det(\bm F'\bm A\bm F+\rho \bm R_i^{-1})=\log\det(\sum_{i=1}^n a_i\bm f(\bm x_i)\bm f(\bm x_i)'+\rho \bm R_i^{-1})$ with $\bm A=\diag\{a_1, \ldots, a_n\}$ is a concave function of $\bm a=(a_1, \ldots, a_n)'$ for $\bm a\in [0, 1]^n$.
Denote $\bm B=\diag\{b_1,\ldots, b_n\}$ for $\bm b=(b_1,\ldots,b_n)'\in [0, 1]^n$ and $\bm K=\bm F'\bm A\bm F+\rho\bm R_i^{-1}$.
Define $g(t)$ as follows.
\begin{align*}
g(t)&=\log\det\left(\bm F'\bm A\bm F+\rho \bm R_i^{-1}+\bm F'(t\bm B)\bm F+\rho \bm R_i^{-1}\right)\\
&=\log\det(\bm F'\bm A\bm F+\rho\bm R_i^{-1})\\
&+\log\det(\bm I_q+(\bm F'\bm A\bm F+\rho \bm R_i^{-1})^{-1/2}(\bm F't\bm B\bm F+\rho \bm R_i^{-1})(\bm F'\bm A\bm F+\rho \bm R_i^{-1})^{-1/2})\\
&=\log\det(\bm F'\bm A\bm F+\rho\bm R_i^{-1})\\
&+\log\det(\bm I_q+t\bm K^{-1/2}\bm F'\bm B\bm F\bm K^{-1/2}+\rho\bm K^{-1/2}\bm R_i^{-1}\bm K^{-1/2})\\
&=\log\det(\bm F'\bm A\bm F+\rho\bm R_i^{-1})+\log\det(\bm I_q+\rho\bm K^{-1/2}\bm R_i^{-1}\bm K^{-1/2})+\sum_{i=1}^q\log(1+t\lambda_i),
\end{align*}
where $\lambda_i$'s are the eigenvalues of
\[(\bm I_q+\rho\bm K^{-1/2}\bm R_i^{-1}\bm K^{-1/2})^{-1/2}\bm K^{-1/2}\bm F'\bm B\bm F\bm K^{-1/2}(\bm I_q+\rho\bm K^{-1/2}\bm R_i^{-1}\bm K^{-1/2})^{-1/2}.\]
Therefore, $\lambda_i\geq 0$ for $i=1, \ldots, q$.
Thus $g(t)$ is a concave function in $t$ for any choice of $\bm a\in [0,1]^n$, which is the sufficient and necessary condition for $\log\det(\bm F'\bm A\bm F+\rho \bm R_i^{-1})$ to be a concave function of $\bm a$.
\end{proof}

\noindent{\bf Proof of the deletion function \eqref{eq:deletion}}
\begin{proof} Denote $\bm F_{-i}$ as the model matrix for $\bm X_{-i}$, which is the design matrix without the $i$th design point, and $\bm W_{0,-i}$, $\bm W_{1,-i}$, and $\bm W_{2,-i}$ the weight matrices accordingly.
According to the properties of matrix determinants, we can show the following.
\begin{align*}
\det\left(\bm F'_{-i}\bm W_{0, -i}\bm F_{-i}\right)&=\det\left(\sum_{j\neq i}\pi(\bm x_i, \bm \eta)(1-\pi(\bm x_i, \bm \eta))\bm f(\bm x_j)\bm f(\bm x_j)'\right)\\
&=\det\left(\bm F'\bm W_0\bm F-\pi(\bm x_i, \bm \eta)(1-\pi(\bm x_i, \bm \eta))\bm f(\bm x_i)\bm f(\bm x_i)'\right)\\
&=\det(\bm F'\bm W_0\bm F)\left[1-\pi(\bm x_i, \bm \eta)(1-\pi(\bm x_i, \bm \eta))\bm f(\bm x_i)'(\bm F'\bm W_0\bm F)^{-1}\bm f(\bm x_i)\right]\\
&=\det(\bm F'\bm W_0\bm F)\left[1-\pi(\bm x_i, \bm \eta)(1-\pi(\bm x_i, \bm \eta))v_0(\bm x_i)\right].
\end{align*}
\begin{align*}
&\det\left(\bm F'_{-i}\bm W_{1, -i}\bm F_{-i}+\rho\bm R^{-1}\right)\\
=&\det\left(\sum_{j\neq i}\pi(\bm x_i, \bm \eta)\bm f(\bm x_j)\bm f(\bm x_j)'+\rho\bm R^{-1}\right)\\
=&\det\left(\bm F'\bm W_1\bm F+\rho\bm R^{-1}-\pi(\bm x_i, \bm \eta)\bm f(\bm x_i)\bm f(\bm x_i)'\right)\\
=&\det\left(\bm F'\bm W_1\bm F+\rho\bm R^{-1}\right)\left[1-\pi(\bm x_i, \bm \eta)\bm f(\bm x_i)'(\bm F'\bm W_1\bm F+\rho\bm R^{-1})^{-1}\bm f(\bm x_i)\right]\\
=&\det\left(\bm F'\bm W_1\bm F+\rho\bm R^{-1}\right)\left[1-\pi(\bm x_i, \bm \eta)v_1(\bm x_i)\right].
\end{align*}
Similarly,
\[
\det\left(\bm F'_{-i}\bm W_{2, -i}\bm F_{-i}+\rho\bm R_2^{-1}\right)=\det\left(\bm F'\bm W_2\bm F+\rho\bm R^{-1}\right)\left[1-(1-\pi(\bm x_i, \bm \eta))v_2(\bm x_i)\right].
\]
The deletion function \eqref{eq:deletion} is then obtained.
\end{proof}

\noindent{\bf Shortcut formulas for $\bm M_{i,-j}$.}
\begin{proof}
Denote $\bm M_{i,-j}$ for $i=0, 1, 2$ as the updated $\bm M_i$ when the $j$th design point is removed.
The following shortcut formulas are used in constructing the initial design.
\[
\bm M_{i,-j}=\bm M_i+\left[\frac{\pi(\bm x_j,\bm\eta)(1-\pi(\bm x_j,\bm\eta))}{1-\pi(\bm x_j,\bm\eta)(1-\pi(\bm x_j,\bm\eta))v_0(\bm x_j)}\right]\bm M_i\bm f(\bm x_j)\bm f(\bm x_j)'\bm M_i, \quad\textrm{ for }i=0,1,2.
\]
\end{proof}

\noindent{\bf Proof of $\Delta(\bm x, \bm x_i)$ in \eqref{eq:delta}.}
\begin{proof}
Denote $\bm F^*$ as the updated model matrix for updated design $\bm X^*$, and $\bm W_i^*$ for $i=0, 1, 2$ as the updated weight matrices accordingly.
Let
\[\bm G_0=[\sqrt{\pi(\bm x, \bm \eta)(1-\pi(\bm x, \bm \eta)}\bm f(\bm x), i\sqrt{\pi(\bm x_i, \bm \eta)(1-\pi(\bm x_i, \bm \eta)}\bm f(\bm x_i)],
\]
where $i=\sqrt{-1}$.
\begin{align*}
&\det(\bm F^{*'}\bm W_0^*\bm F^*)=\det\left(\sum_{j\neq i}\pi(\bm x_j, \bm \eta)(1-\pi(\bm x_j, \bm \eta))\bm f(\bm x_j)\bm f(\bm x_j)'\right.\\
&\left.-\pi(\bm x_i, \bm \eta)(1-\pi(\bm x_i, \bm \eta))\bm f(\bm x_i)\bm f(\bm x_i)'+\pi(\bm x, \bm \eta)(1-\pi(\bm x, \bm \eta))\bm f(\bm x)\bm f(\bm x)'\right)\\
&=\det(\bm F'\bm W_0\bm F+\bm G_0\bm G_0')\\
&=\det(\bm F'\bm W_0\bm F)\det\left(\bm I_2+\bm G_0'\bm M_0\bm G_0\right).
\end{align*}
It can be computed that
\[\det\left(\bm I_2+\bm G_0'\bm M_0\bm G_0\right)=\Delta_0(\bm x, \bm x_i), \]
where $\Delta_0(\bm x, \bm x_i)$ is
\begin{align*}
\Delta_0(\bm x, \bm x_i)&=\left[1+\pi(\bm x, \bm \eta)(1-\pi(\bm x, \bm \eta))v_0(\bm x)\right]\left[1-\pi(\bm x_i, \bm \eta)(1-\pi(\bm x_i, \bm \eta))v_0(\bm x_i)\right]\\\nonumber
&+\pi(\bm x, \bm \eta)(1-\pi(\bm x, \bm \eta))\pi(\bm x_i, \bm \eta)(1-\pi(\bm x_i, \bm \eta))v_0(\bm x,\bm x_i)^2.
\end{align*}
Similarly, define
\begin{align*}
\bm G_1&=[\sqrt{\pi(\bm x, \eta)}\bm f(\bm x), i\sqrt{\pi(\bm x_i, \eta)}\bm f(\bm x_i)]\quad\textrm{and}\\
\bm G_2 &=[\sqrt{(1-\pi(\bm x, \bm \eta)}\bm f(\bm x), i\sqrt{(1-\pi(\bm x_i, \bm \eta)}\bm f(\bm x_i)].
\end{align*}
Following a similar derivation,
\begin{align*}
\det(\bm F^{*'}\bm W_1^*\bm F^*+\rho\bm R)&=\det(\bm F'\bm W_1\bm F+\rho\bm R)\Delta_1(\bm x, \bm x_i),\\
\det(\bm F^{*'}\bm W_2^*\bm F^*+\rho\bm R)&=\det(\bm F'\bm W_2\bm F+\rho\bm R)\Delta_2(\bm x, \bm x_i),
\end{align*}
where $\Delta_1(\bm x, \bm x_i)$ and $\Delta_2(\bm x, \bm x_i)$ are
\begin{align*}
\Delta_1(\bm x, \bm x_i)&=\left[1+\pi(\bm x,\bm \eta)v_1(\bm x)\right]\left[1-\pi(\bm x_i,\bm \eta)v_1(\bm x_i)\right]+\pi(\bm x,\bm \eta)\pi(\bm x_i,\bm \eta)v_1(\bm x, \bm x_i)^2,\\
\Delta_2(\bm x, \bm x_i)&=\left[1+(1-\pi(\bm x,\bm \eta))v_2(\bm x)\right]\left[1-(1-\pi(\bm x_i,\bm \eta))v_2(\bm x_i)\right]\\
&+(1-\pi(\bm x,\bm \eta))(1-\pi(\bm x_i,\bm \eta))v_2(\bm x, \bm x_i)^2.
\end{align*}
Thus $\Delta(\bm x, \bm x_i)$ is computed as in \eqref{eq:delta}.
\end{proof}

\noindent{\bf Proof of update formulas $\bm M_i^*$ for $i=0, 1, 2$}
\begin{proof}
Use the same notation of $\bm G_i$ for $i=0, 1, 2$ as in the previous proof.
Define the functions
\begin{align*}
&\bm S_i(\bm x)=\bm M_i\bm f(\bm x)\bm f(\bm x)'\bm M_i, \quad\textrm{for } i=0, 1, 2,\\
&\bm S_i(\bm x_1, \bm x_2)=\bm M_i\bm f(\bm x_1)\bm f(\bm x_2)'\bm M_i, \quad\textrm{for } i=0, 1, 2.
\end{align*}
It is straightforward to derive
\begin{align*}
\bm M_0^*&=\left(\bm F'\bm W_0\bm F+\bm G_0\bm G_0'\right)^{-1}\\
&=\bm M_0-\bm M_0 \bm G_0 \left(\bm I_2+\bm G_0'\bm M_0\bm G_0\right)^{-1}\bm G'\bm M_0.
\end{align*}
For simpler notation, denote $a(\bm x)=\pi(\bm x, \bm \eta)(1-\pi(\bm x, \bm \eta))$ and $a(\bm x_i)=\pi(\bm x_i, \bm \eta)(1-\pi(\bm x_i, \bm \eta))$.
\begin{align*}
&\left(\bm I_2+\bm G_0'\bm M_0\bm G_0\right)^{-1}\\
=&\left(
\begin{array}{ll}
1+a(\bm x)v_0(\bm x),& i\sqrt{a(\bm x)a(\bm x_i)}v_0(\bm x, \bm x_i)\\
 i\sqrt{a(\bm x)a(\bm x_i)}v_0(\bm x, \bm x_i), & 1-a(\bm x_i)v_0(\bm x_i)
\end{array}
\right)^{-1}\\
=&\frac{1}{\Delta_0(\bm x,\bm x_i)}\left(
\begin{array}{ll}
1-a(\bm x_i)v_0(\bm x_i),& -i\sqrt{a(\bm x)a(\bm x_i)}v_0(\bm x, \bm x_i)\\
 -i\sqrt{a(\bm x)a(\bm x_i)}v_0(\bm x, \bm x_i), & 1+a(\bm x)v_0(\bm x)
\end{array}
\right).
\end{align*}

\begin{align*}
\bm M_0^*&=\bm M_0-\frac{1}{\Delta_0(\bm x, \bm x_i)}\bm M_0\left[\sqrt{a(\bm x)}v_0(\bm x), i\sqrt{a(\bm x_i)}v_0(\bm x_i)\right]\\
\times &\left(
\begin{array}{ll}
1-a(\bm x_i)v_0(\bm x_i),& -i\sqrt{a(\bm x)a(\bm x_i)}v_0(\bm x, \bm x_i)\\
 -i\sqrt{a(\bm x)a(\bm x_i)}v_0(\bm x, \bm x_i), & 1+a(\bm x)v_0(\bm x)
\end{array}
\right)
\times
\left[
\begin{array}{c}
\sqrt{a(\bm x)}\bm f(\bm x)'\\
i\sqrt{a_i(\bm x_i)}\bm f(\bm x_i)'
\end{array}
\right]\bm M_0.\\
&=\bm M_0-\frac{1}{\Delta_0(\bm x, \bm x_i)}\left\{\pi(\bm x, \bm \eta)(1-\pi(\bm x, \bm \eta))\left[1-\pi(\bm x_i, \bm \eta)(1-\pi(\bm x_i, \bm \eta))v_0(\bm x_i)\right]\bm S_0(\bm x)\right.\\
&\left.-\pi(\bm x_i, \bm \eta)(1-\pi(\bm x_i, \bm \eta))\left[1+\pi(\bm x, \bm \eta)(1-\pi(\bm x, \bm \eta))v_0(\bm x)\right]\bm S_0(\bm x_i)\right.\\
&+\left.\pi(\bm x_i, \bm \eta)(1-\pi(\bm x_i, \bm \eta))\pi(\bm x, \bm \eta)(1-\pi(\bm x, \bm \eta))v_0(\bm x, \bm x_i)\left[\bm S_0(\bm x, \bm x_i)+\bm S_0(\bm x_i, \bm x)\right]\right\}
\end{align*}
Update formulas for $\bm M_1^*$ and $\bm M_2^*$ can be derived similarly as follows.
\begin{align*}
\bm M_1^*=\bm M_1&-\frac{1}{\Delta_1(\bm x, \bm x_i)}\left\{\pi(\bm x, \bm \eta)\left[1-\pi(\bm x_i, \bm \eta)v_1(\bm x_i)\right]\bm S_1(\bm x)-\pi(\bm x_i, \bm \eta)\left[1+\pi(\bm x, \bm \eta)v_1(\bm x)\right]\bm S_1(\bm x_i)\right.\\
&+\left.\pi(\bm x_i, \bm \eta)\pi(\bm x, \bm \eta)v_1(\bm x, \bm x_i)\left[\bm S_1(\bm x, \bm x_i)+\bm S_1(\bm x_i, \bm x)\right]\right\},\\
\bm M_2^*=\bm M_2&-\frac{1}{\Delta_2(\bm x, \bm x_i)}\left\{(1-\pi(\bm x, \bm \eta))\left[1-(1-\pi(\bm x_i, \bm \eta))v_2(\bm x_i)\right]\bm S_2(\bm x)\right.\\
&\left.-(1-\pi(\bm x_i, \bm \eta))\left[1+(1-\pi(\bm x, \bm \eta))v_2(\bm x)\right]\bm S_2(\bm x_i)\right.\\
&+\left.(1-\pi(\bm x_i, \bm \eta))(1-\pi(\bm x, \bm \eta))v_2(\bm x, \bm x_i)\left[\bm S_2(\bm x, \bm x_i)+\bm S_2(\bm x_i, \bm x)\right]\right\}.
\end{align*}
\end{proof}

\section{Appendix A2. Table}\label{sec:table}

\begin{table}
\begin{center}
\caption{Local Bayesian $D$-optimal Designs for $\rho=0$ and 0.3 and other three alternative designs. 
The values in columns 7-11 are frequencies. \label{tab:arti-fix}}
\begin{tiny}
\begin{tabular}{|c|ccccc|c|c|c|c|c|}\hline
Point & $x_1$ & $x_2$ & $x_3$ & $x_4$ & $x_5$ & $D_{QQ}$ $\rho=0$ & $D_{QQ}$ $\rho=0.3$ & $D_G$ & $D_L$ & $D_n$ \\\hline
1&-1&-1&-1&-1&-1&1&2&1&1&2\\
2&1&-1&-1&-1&-1&1&1&2&1&1\\
3&-1&1&-1&-1&-1&1&1&2&1&1\\
4&1&1&-1&-1&-1&2&1&2&1&2\\
5&-1&-1&1&-1&-1&1&0&2&1&1\\
6&1&-1&1&-1&-1&1&1&0&1&1\\
7&-1&1&1&-1&-1&2&2&1&1&2\\
8&1&1&1&-1&-1&1&2&1&1&1\\
9&-1&-1&-1&0&-1&2&2&1&1&1\\
10&1&-1&-1&0&-1&0&0&1&1&1\\
11&-1&1&-1&0&-1&1&1&2&1&2\\
12&1&1&-1&0&-1&1&2&1&1&1\\
13&-1&-1&1&0&-1&1&1&1&1&2\\
14&1&-1&1&0&-1&2&2&1&1&1\\
15&-1&1&1&0&-1&1&1&0&1&0\\
16&1&1&1&0&-1&1&1&1&1&2\\
17&-1&-1&-1&1&-1&1&0&1&1&2\\
18&1&-1&-1&1&-1&2&2&1&1&1\\
19&-1&1&-1&1&-1&2&2&1&1&1\\
20&1&1&-1&1&-1&1&1&1&1&1\\
21&-1&-1&1&1&-1&2&2&1&1&1\\
22&1&-1&1&1&-1&0&1&1&1&1\\
23&-1&1&1&1&-1&0&1&1&1&2\\
24&1&1&1&1&-1&2&1&1&1&1\\
25&-1&-1&-1&-1&0&1&0&1&1&0\\
26&1&-1&-1&-1&0&0&0&1&1&1\\
27&-1&1&-1&-1&0&0&0&0&1&0\\
28&1&1&-1&-1&0&0&1&1&1&0\\
29&-1&-1&1&-1&0&0&1&1&1&1\\
30&1&-1&1&-1&0&0&0&0&1&0\\
31&-1&1&1&-1&0&0&0&0&1&0\\
32&1&1&1&-1&0&1&0&1&1&0\\
33&-1&-1&-1&0&0&0&0&1&1&0\\
34&1&-1&-1&0&0&1&1&1&1&1\\
35&-1&1&-1&0&0&0&0&0&1&0\\
36&1&1&-1&0&0&1&0&1&1&0\\
37&-1&-1&1&0&0&1&1&1&1&0\\
38&1&-1&1&0&0&0&0&1&1&0\\
39&-1&1&1&0&0&0&0&0&1&0\\
40&1&1&1&0&0&0&0&1&1&0\\
41&-1&-1&-1&1&0&0&2&1&0&0\\
42&1&-1&-1&1&0&0&0&1&1&0\\
43&-1&1&-1&1&0&0&0&0&1&0\\
44&1&1&-1&1&0&1&0&1&0&0\\
45&-1&-1&1&1&0&0&0&1&1&0\\
46&1&-1&1&1&0&1&1&1&0&0\\
47&-1&1&1&1&0&1&0&1&0&0\\
48&1&1&1&1&0&0&1&1&1&1\\
49&-1&-1&-1&-1&1&1&1&1&1&1\\
50&1&-1&-1&-1&1&2&2&1&1&2\\
51&-1&1&-1&-1&1&1&1&0&1&1\\
52&1&1&-1&-1&1&1&1&1&1&1\\
53&-1&-1&1&-1&1&2&1&1&1&2\\
54&1&-1&1&-1&1&0&0&0&1&0\\
55&-1&1&1&-1&1&1&1&1&1&1\\
56&1&1&1&-1&1&1&2&1&1&2\\
57&-1&-1&-1&0&1&1&1&1&1&2\\
58&1&-1&-1&0&1&1&1&1&1&0\\
59&-1&1&-1&0&1&2&2&1&1&1\\
60&1&1&-1&0&1&1&1&1&1&2\\
61&-1&-1&1&0&1&2&2&1&0&1\\
62&1&-1&1&0&1&1&1&1&1&2\\
63&-1&1&1&0&1&0&0&1&1&1\\
64&1&1&1&0&1&2&2&1&0&1\\
65&-1&-1&-1&1&1&2&1&1&1&2\\
66&1&-1&-1&1&1&1&1&1&1&1\\
67&-1&1&-1&1&1&1&1&1&1&1\\
68&1&1&-1&1&1&1&2&1&1&2\\
69&-1&-1&1&1&1&1&1&1&1&1\\
70&1&-1&1&1&1&1&1&0&1&1\\
71&-1&1&1&1&1&2&2&1&1&2\\
72&1&1&1&1&1&1&0&1&1&1\\\hline
\end{tabular}
\end{tiny}
\end{center}
\end{table}

\end{document}